\PassOptionsToPackage{unicode}{hyperref}
\PassOptionsToPackage{hyphens}{url}
\documentclass[
]{article}
\usepackage{amsmath,amssymb}
\usepackage{lmodern}
\usepackage{iftex}
\ifPDFTeX
  \usepackage[T1]{fontenc}
  \usepackage[utf8]{inputenc}
  \usepackage{textcomp} 
\else 
  \usepackage{unicode-math}
  \defaultfontfeatures{Scale=MatchLowercase}
  \defaultfontfeatures[\rmfamily]{Ligatures=TeX,Scale=1}
\fi
\IfFileExists{upquote.sty}{\usepackage{upquote}}{}
\IfFileExists{microtype.sty}{
  \usepackage[]{microtype}
  \UseMicrotypeSet[protrusion]{basicmath} 
}{}
\makeatletter
\@ifundefined{KOMAClassName}{
  \IfFileExists{parskip.sty}{%
    \usepackage{parskip}
  }{
    \setlength{\parindent}{0pt}
    \setlength{\parskip}{6pt plus 2pt minus 1pt}}
}{
  \KOMAoptions{parskip=half}}
\makeatother
\usepackage{xcolor}
\IfFileExists{xurl.sty}{\usepackage{xurl}}{} 
\IfFileExists{bookmark.sty}{\usepackage{bookmark}}{\usepackage{hyperref}}
\hypersetup{
  pdftitle={Finite-sample bias-correction factors for the median absolute deviation based on the Harrell-Davis quantile estimator and its trimmed modification},
  hidelinks,
  pdfcreator={LaTeX via pandoc}}
\usepackage[margin=1in]{geometry}
\usepackage{color}
\usepackage{fancyvrb}

\DefineVerbatimEnvironment{Highlighting}{Verbatim}{commandchars=\\\{\}}
\usepackage{framed}
\definecolor{shadecolor}{RGB}{248,248,248}
\newenvironment{Shaded}{\begin{snugshade}}{\end{snugshade}}

\newcommand{\AttributeTok}[1]{\textcolor[rgb]{0.77,0.63,0.00}{#1}}

\newcommand{\CommentTok}[1]{\textcolor[rgb]{0.56,0.35,0.01}{\textit{#1}}}

\newcommand{\ConstantTok}[1]{\textcolor[rgb]{0.00,0.00,0.00}{#1}}
\newcommand{\ControlFlowTok}[1]{\textcolor[rgb]{0.13,0.29,0.53}{\textbf{#1}}}

\newcommand{\DecValTok}[1]{\textcolor[rgb]{0.00,0.00,0.81}{#1}}

\newcommand{\FloatTok}[1]{\textcolor[rgb]{0.00,0.00,0.81}{#1}}
\newcommand{\FunctionTok}[1]{\textcolor[rgb]{0.00,0.00,0.00}{#1}}

\newcommand{\NormalTok}[1]{#1}

\newcommand{\OtherTok}[1]{\textcolor[rgb]{0.56,0.35,0.01}{#1}}

\newcommand{\SpecialCharTok}[1]{\textcolor[rgb]{0.00,0.00,0.00}{#1}}

\usepackage{longtable,booktabs,array}
\usepackage{calc} 
\usepackage{etoolbox}
\makeatletter
\patchcmd\longtable{\par}{\if@noskipsec\mbox{}\fi\par}{}{}
\makeatother
\IfFileExists{footnotehyper.sty}{\usepackage{footnotehyper}}{\usepackage{footnote}}
\makesavenoteenv{longtable}
\usepackage{graphicx}
\makeatletter
\def\maxwidth{\ifdim\Gin@nat@width>\linewidth\linewidth\else\Gin@nat@width\fi}
\def\maxheight{\ifdim\Gin@nat@height>\textheight\textheight\else\Gin@nat@height\fi}
\makeatother
\setkeys{Gin}{width=\maxwidth,height=\maxheight,keepaspectratio}
\makeatletter
\def\fps@figure{htbp}
\makeatother
\providecommand{\tightlist}{%
  \setlength{\itemsep}{0pt}\setlength{\parskip}{0pt}}
\usepackage{hyperref}
\definecolor{linkcolor}{HTML}{D55E00}
\definecolor{citecolor}{HTML}{009E73}
\definecolor{urlcolor}{HTML}{0072B2}
\hypersetup{
    colorlinks,
    linkcolor={linkcolor},
    citecolor={citecolor},
    urlcolor={urlcolor}
}

\renewcommand{\DecValTok}[1]{\textcolor[HTML]{009E73}{#1}}
\renewcommand{\FloatTok}[1]{\textcolor[HTML]{009E73}{#1}}
\renewcommand{\ConstantTok}[1]{\textcolor[HTML]{009E73}{#1}}
\renewcommand{\ControlFlowTok}[1]{\textcolor[HTML]{0072B2}{\textbf{#1}}}
\renewcommand{\OtherTok}[1]{\textcolor[HTML]{000000}{#1}}
\renewcommand{\CommentTok}[1]{\textcolor[HTML]{999999}{\textit{#1}}}
\renewcommand{\AttributeTok}[1]{\textcolor[HTML]{CC79A7}{#1}}
\renewcommand{\FunctionTok}[1]{\textcolor[HTML]{56B4E9}{#1}}

\usepackage{caption}
\usepackage[vlined]{algorithm2e}
\usepackage{algorithmic}

\usepackage{titlesec}
\titlespacing{\section}{0pt}{\parskip}{0pt}
\titlespacing{\subsection}{0pt}{\parskip}{0pt}

\usepackage{enumitem}
\setlist[itemize]{topsep=0pt}
\usepackage{amsmath}
\usepackage{float}
\usepackage{booktabs}
\usepackage{longtable}
\usepackage{array}
\usepackage{multirow}
\usepackage{wrapfig}
\usepackage{colortbl}
\usepackage{pdflscape}
\usepackage{tabu}
\usepackage{threeparttable}
\usepackage{threeparttablex}
\usepackage[normalem]{ulem}
\usepackage{makecell}
\usepackage{xcolor}
\ifLuaTeX
  \usepackage{selnolig}  
\fi
\usepackage[style=alphabetic,sorting=nty]{biblatex}
\title{Finite-sample bias-correction factors for the median absolute deviation based on the Harrell-Davis quantile estimator and its trimmed modification}
\author{Andrey Akinshin\\
Huawei Research, \href{mailto:andrey.akinshin@gmail.com}{\nolinkurl{andrey.akinshin@gmail.com}}}
\date{}

\begin{document}
\maketitle
\begin{abstract}
The median absolute deviation is a widely used robust measure of statistical dispersion.
Using a scale constant, we can use it as an asymptotically consistent estimator for the standard deviation under normality.
For finite samples, the scale constant should be corrected in order to obtain an unbiased estimator.
The bias-correction factor depends on the sample size and the median estimator.
When we use the traditional sample median, the factor values are well known,
but this approach does not provide optimal statistical efficiency.
In this paper, we present the bias-correction factors for the median absolute deviation
based on the Harrell-Davis quantile estimator and its trimmed modification
which allow us to achieve better statistical efficiency of the standard deviation estimations.
The obtained estimators are especially useful for samples with a small number of elements.

\end{abstract}

\hypertarget{introduction}{%
\section{Introduction}\label{introduction}}

We consider the median absolute deviation as a robust alternative to the standard deviation.
In order to make it asymptotically consistent with the standard deviation under the normal distribution,
the median absolute deviation should be multiplied by a scale constant \(C_\infty \approx 1.4826\).
This approach works well in practice when the sample size \(n\) is large.
However, when the sample size is small, the usage of \(C_\infty\) produces a biased estimator.
The goal of this paper is to provide proper bias-correction factors \(C_n\) for finite samples.

When the median absolute deviation is based on the traditional sample median, these factors are known (see \autocite{park2020}).
However, the sample median is not the most statistically efficient way to estimate the true population median.
As a more efficient alternative, we can use the Harrell-Davis quantile estimator (see \autocite{harrell1982})
to calculate the median absolute deviation.
This approach is more efficient than the classic sample median, but it is not robust.
To achieve a trade-off between statistical efficiency and robustness,
we consider a trimmed modification of the Harrell-Davis quantile estimator (see \autocite{akinshin2022}).
The bias-correction factors depend not only on the sample size but also on the chosen median estimator.
Therefore, if we want to use the mentioned quantile estimators, we need adjusted factor values.

In this paper, we present finite-sample bias-correction factors for the median absolute deviation based on
the Harrell-Davis quantile estimator and its trimmed modifications.
For \(n=2\), we derive the exact factor value.
For \(3\leq n \leq 100\), we obtain factor values using Monte-Carlo simulations.
For \(n>100\), we provide a prediction equation using the least squares method.

The suggested approach provides a robust estimator of the standard deviation
that is unbiased under normality and more efficient than the classic approach based on the sample median.

The paper is organized as follows.
In Section~\protect\hyperlink{sec-preliminaries}{2},
we introduce the preliminaries with the problem explanation, a historical overview, and relevant references.
In Section~\protect\hyperlink{sec-simulations}{3}, we perform a series of numerical simulations
to get the values of the bias-correction factors and analyze properties of the obtained estimators.
In Section~\protect\hyperlink{sec-cases}{4},
we get the exact bias-correction factor value for \(n=2\) and a prediction equation for \(n>100\).
In Section~\protect\hyperlink{sec-summary}{5}, we summarize all the results.
In Appendix~\protect\hyperlink{sec-ref}{A}, we provide a reference R implementation of the presented unbiased estimators.

\clearpage

\hypertarget{sec-preliminaries}{%
\section{Preliminaries}\label{sec-preliminaries}}

In this section, we provide the motivation for the present research, a historical overview of the subject,
relevant background and references.

\hypertarget{measures-of-statistical-dispersion}{%
\subsection{Measures of statistical dispersion}\label{measures-of-statistical-dispersion}}

\begin{figure}[ht!]

{\centering \includegraphics{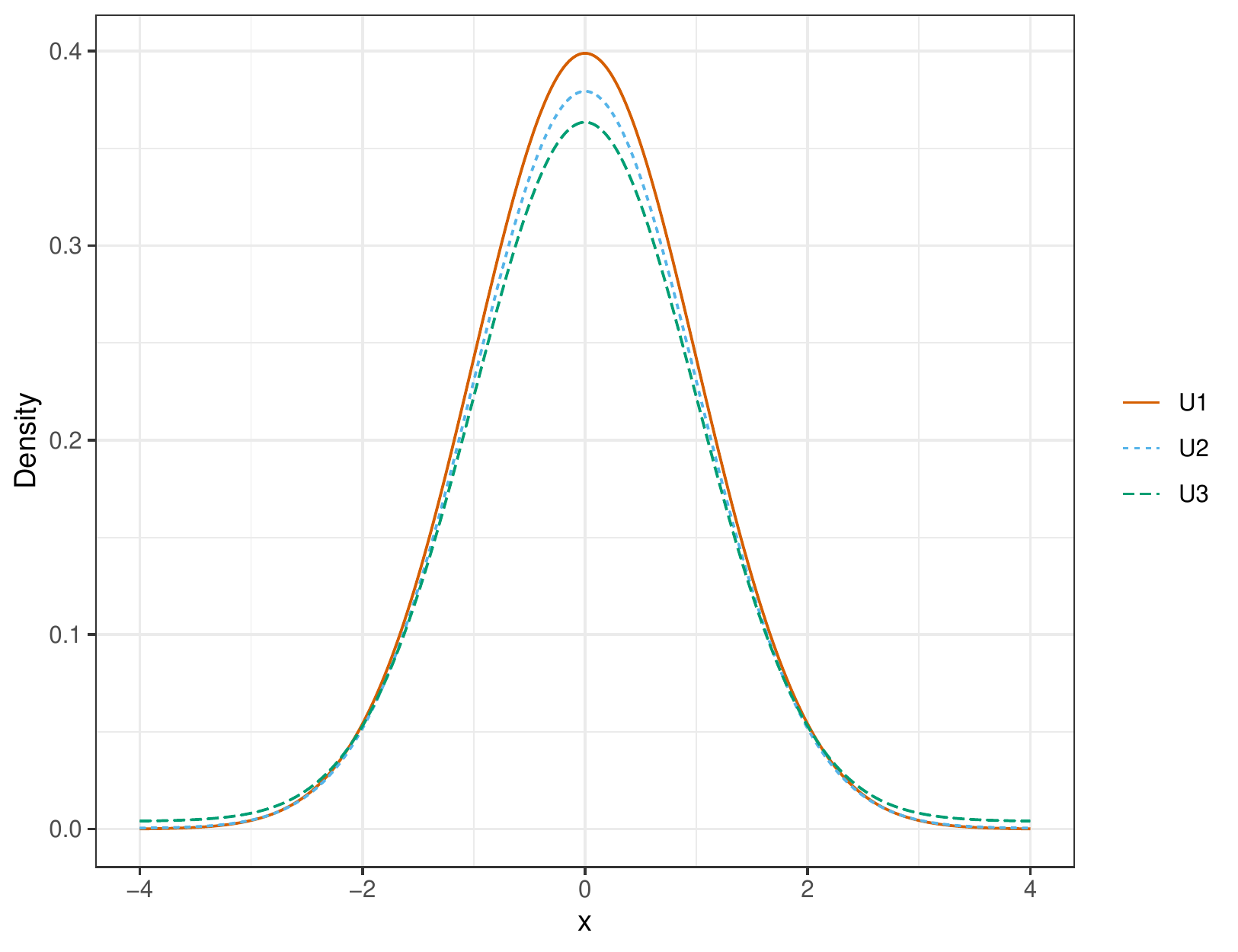} 

}

\caption{Three density plots of close-to-normal distributions.}\label{fig:stddev1}
\end{figure}

The most popular measure of statistical dispersion is the standard deviation.
The classic equations for the standard deviation work great for samples from the normal distributions.
Unfortunately, the real-life experimental data are often contaminated by outliers.
The standard deviation is too sensitive to distribution tails and sample outliers
so that it can be easily corrupted by a single extreme value.
For example, let us consider three density plots presented in Figure~\ref{fig:stddev1}.

\begin{figure}[ht!]

{\centering \includegraphics{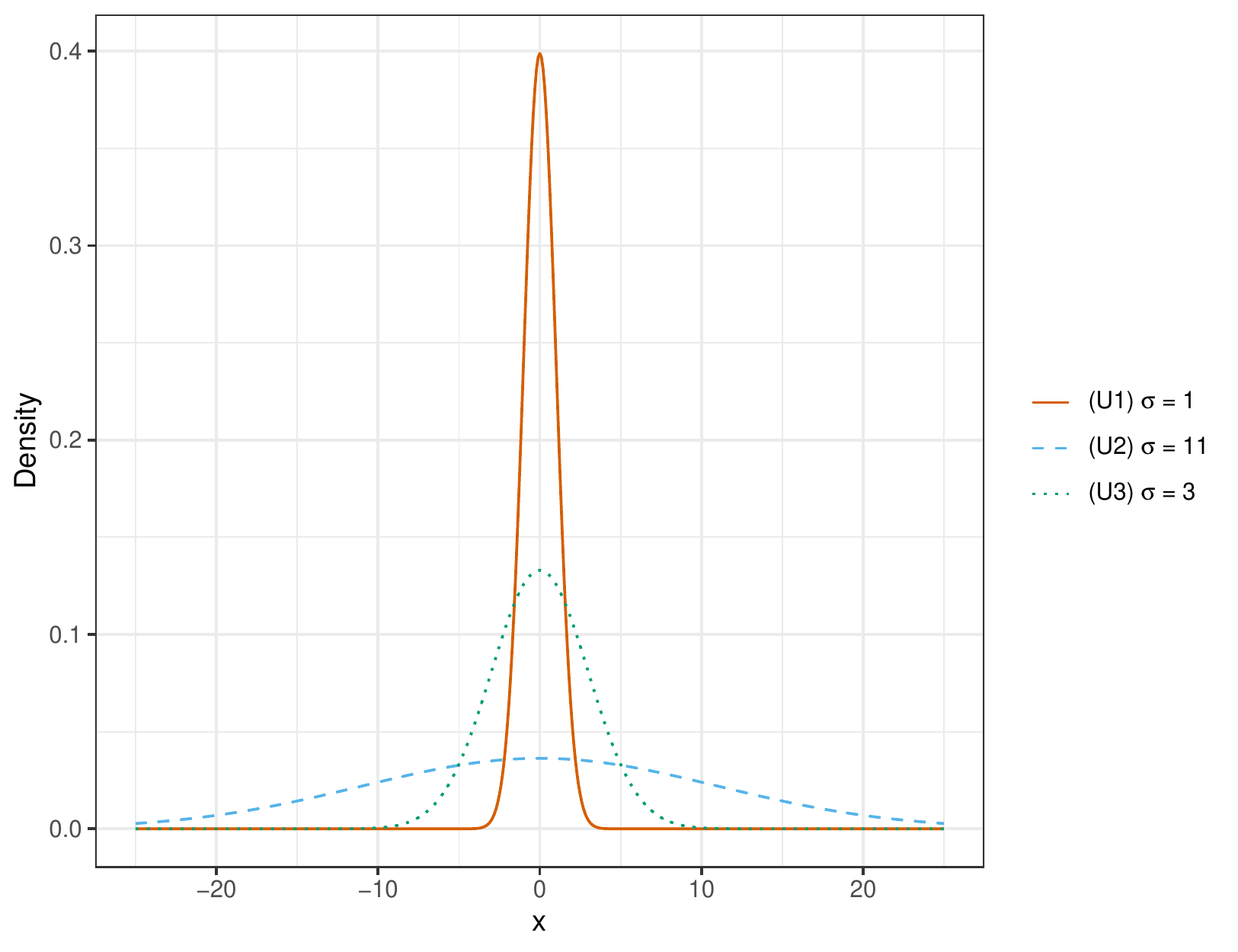} 

}

\caption{Normal distributions with various standard deviation values.}\label{fig:stddev2}
\end{figure}

All three presented distributions look quite close to the normal one.
However, their actual standard deviations are \(\sigma_{U1} = 1\), \(\sigma_{U2} = 11\), \(\sigma_{U3} = 3\).
The actual normal distributions with such standard deviations are presented in Figure~\ref{fig:stddev2}.

In fact, only the first distribution is the standard normal distribution \(U1 = \mathcal{N}(0, 1^2)\).
Two other distributions are contaminated normal distributions
that are mixtures of two normal distributions (see \autocite{wilcox2016}):
\(U2 = 0.95\cdot\mathcal{N}(0, 1^2) + 0.05\cdot\mathcal{N}(0, 49^2)\),
\(U3 = 0.9\cdot\mathcal{N}(0, 1^2) + 0.1\cdot\mathcal{N}(0, 9^2)\).
The standard deviation is not a robust measure, its breakdown point is zero.
If the data are contaminated, the standard deviation estimations can be misleading
because of the high sensitivity to outliers.

In order to solve this problem, we may need a robust measure of the statistical dispersion.
A widely used option is the \emph{median absolute deviation}.
One of the first mentions can be found in \autocite{hampel1974} where it is attributed to Gauss.
In \autocite{rousseeuw1993}, the median absolute deviation is introduced as a very robust scale estimator
because it has the best possible breakdown point (0.5).

Let \(X\) be a sample of i.i.d. random variables: \(X = \{ X_1, X_2, \ldots X_n \}\).
Here is the classic non-scaled definition of the median absolute deviation:

\[
\operatorname{MAD}_0(X) = \operatorname{median}(|X - \operatorname{median}(X)|).
\]

Let us assume that \(X\) follows the standard normal distribution: \(X \sim \mathcal{N}(0,1^2)\).
If we want to make the median absolute deviation asymptotically consistent
with the standard deviation under the normal distribution
for an infinitely large sample,
we should define a modification of \(\operatorname{MAD}_0\) with a bias-correction factor \(C_\infty\):

\[
\operatorname{MAD}_\infty(X) = C_\infty \cdot \operatorname{median}(|X - \operatorname{median}(X)|) = C_\infty \cdot \operatorname{MAD}_0(X).
\]

Since we are building an unbiased estimator,
its asymptotic expected value \(\lim_{n\to\infty}\mathbb{E}[\operatorname{MAD}_\infty(X)]\) should be equal to \(1\).
It gives us the following equation for \(C_\infty\):

\[
C_\infty = \frac{1}{\lim_{n\to\infty}\mathbb{E}[\operatorname{MAD}_0(X)]}.
\]

Let us denote \(\lim_{n\to\infty}\mathbb{E}[\operatorname{MAD}_0(|X|)]\) by \(M_\infty\).
Since the median of \(\mathcal{N}(0,1^2)\) is zero, we have:

\[
M_\infty =
  \lim_{n\to\infty}\mathbb{E}[\operatorname{median}(|X - \operatorname{median}(X))] = 
  \lim_{n\to\infty}\mathbb{E}[\operatorname{median}(|X|)].
\]

Since \(M_\infty\) is the expected value of the median of \(|X|\), we can write

\[
\mathbb{P}(|X_1|<M_\infty) = 0.5,
\]

which is the same as

\[
\mathbb{P}(-M_\infty<X_1<M_\infty) = 0.5.
\]

Let us denote the cumulative distribution function of \(\mathcal{N}(0, 1^2)\) by \(\Phi\).
Then, the probability of getting \(X_1\) from the range \((-M_\infty;M_\infty)\)
is \(\Phi(M_\infty)-\Phi(-M_\infty)\).
Thus,

\[
\Phi(M_\infty)-\Phi(-M_\infty) = 0.5.
\]

Since \(\mathcal{N}(0, 1^2)\) is symmetric around zero, \(\Phi(-M_\infty)=1-\Phi(M_\infty)\).
Therefore

\[
\Phi(M_\infty) = 0.75.
\]

Assuming that \(\Phi^{-1}\) is the quantile function of \(\mathcal{N}(0, 1^2)\), we have:

\[
M_\infty = \Phi^{-1}(0.75) \approx 0.674489750196082.
\]

Finally,

\[
C_\infty =
  \frac{1}{\lim_{n\to\infty}\mathbb{E}[\operatorname{MAD}_0(X)]} =
  \dfrac{1}{M_\infty} =
  \dfrac{1}{\Phi^{-1}(0.75)} \approx
  1.4826022185056.
\]\\

Now we consider a scaled median absolute deviation
that could be used as an unbiased standard deviation estimator under normality
for a finite sample of size \(n\).
Let us denote it by \(\operatorname{MAD}_n\):

\[
\operatorname{MAD}_n(X) = C_n \cdot \operatorname{median}(|X - \operatorname{median}(X)|) = C_n \cdot \operatorname{MAD}_0(X).
\]

We cannot use \(C_{\infty}\) as a bias-correction factor for finite samples because
it would make \(\operatorname{MAD}_n\) a biased estimator of the standard deviation.
To make it unbiased, we have to find proper values of \(C_n\) for each sample size \(n\).
These values can be evaluated as

\[
C_n = \dfrac{1}{\mathbb{E}[\operatorname{MAD}_0(X)]} = \dfrac{1}{M_n},
\]

where \(M_n = \mathbb{E}[\operatorname{MAD}_0(X)]\),
\(X = \{ X_1, X_2, \ldots, X_n \}\).

\clearpage

\hypertarget{bias-correction-factors-based-on-the-sample-median}{%
\subsection{Bias-correction factors based on the sample median}\label{bias-correction-factors-based-on-the-sample-median}}

Traditionally, by \(\operatorname{median}\) we assume the sample median
(if \(n\) is odd, the median is the middle order statistic;
if \(n\) is even, the median is the arithmetic average of the two middle order statistics).
This approach is consistent with the Hyndman-Fan Type 7 quantile estimator (see \autocite{hyndman1996})
which is the most popular traditional quantile estimator based on one or two order statistics
(it is used by default in R, Julia, NumPy, and Excel).
To avoid confusion, let us denote the median estimator based on the sample median by \(\operatorname{median}_{\operatorname{SM}}\).
Similarly, we denote \(\operatorname{MAD}\) based on \(\operatorname{median}_{\operatorname{SM}}\) by \(\operatorname{MAD}_{\operatorname{SM}}\).
Let us briefly discuss existing approaches for picking \(C_n\) values for \(\operatorname{median}_{\operatorname{SM}}\).

One of the first attempts to define \(C_n\) was made in \autocite{croux1992} by Christophe Croux and Peter J. Rousseeuw.
They suggested using the following equations:

\[
C_n = \dfrac{b_n}{\Phi^{-1}(0.75)}.
\]

For \(n \leq 9\), the approximated values of \(b_n\) were defined as presented in Table~\ref{tab:croux}.

\begin{longtable}[]{@{}rr@{}}
\caption{\label{tab:croux} Original \(b_n\) factors from the Croux-Rousseeuw approach.}\tabularnewline
\toprule
n & \(b_n\) \\
\midrule
\endfirsthead
\toprule
n & \(b_n\) \\
\midrule
\endhead
2 & 1.196 \\
3 & 1.495 \\
4 & 1.363 \\
5 & 1.206 \\
6 & 1.200 \\
7 & 1.140 \\
8 & 1.129 \\
9 & 1.107 \\
\bottomrule
\end{longtable}

For \(n > 9\), they suggested using the following equation:

\[
b_n = \dfrac{n}{n-0.8}.
\]

This approach was improved in \autocite{williams2011} by Dennis C. Williams.
Firstly, he provided updated \(b_n\) values for \(n \leq 9\) (see Table~\ref{tab:williams}).

\begin{longtable}[]{@{}rr@{}}
\caption{\label{tab:williams} Williams version of \(b_n\) factors from the Croux-Rousseeuw approach.}\tabularnewline
\toprule
n & \(b_n\) \\
\midrule
\endfirsthead
\toprule
n & \(b_n\) \\
\midrule
\endhead
2 & 1.197 \\
3 & 1.490 \\
4 & 1.360 \\
5 & 1.217 \\
6 & 1.189 \\
7 & 1.138 \\
8 & 1.127 \\
9 & 1.101 \\
\bottomrule
\end{longtable}

Secondly, he introduced a small correction for \(n > 9\):

\[
b_n = \dfrac{n}{n-0.801}.
\]

Thirdly, he discussed another kind of approximation for such kind of bias-correction factors:

\[
b_n \cong 1 + cn^{-d}.
\]

In his paper, he applied the above equation only to \emph{Shorth}
(which is the smallest interval that contains at least half of the data points),
but this approach can also be applied to other measures of scale.

Next, in \autocite{hayes2014}, Kevin Hayes suggested another kind of prediction equation for \(n \geq 9\):

\[
C_n = \dfrac{1}{\hat{a}_n},
\]

where

\[
\hat{a}_n = \Phi^{-1}(0.75) \Bigg( 1 - \dfrac{\alpha}{n} - \dfrac{\beta}{n^2} \Bigg).
\]

The suggested values of \(\alpha\) and \(\beta\) are listed in Table~\ref{tab:hayes}.

\begin{longtable}[]{@{}rrr@{}}
\caption{\label{tab:hayes} \(\alpha\) and \(\beta\) values from the Hayes approach.}\tabularnewline
\toprule
n & \(\alpha\) & \(\beta\) \\
\midrule
\endfirsthead
\toprule
n & \(\alpha\) & \(\beta\) \\
\midrule
\endhead
odd & 0.7635 & 0.565 \\
even & 0.7612 & 1.123 \\
\bottomrule
\end{longtable}

Finally, in \autocite{park2020}, Chanseok Park, Haewon Kim, and Min Wang aggregated all of the previous results.
They used the following form of the main equation:

\[
C_n = \dfrac{1}{\Phi^{-1}(0.75) \cdot (1+A_n)}.
\]

For \(n > 100\), they suggested two approaches.
The first one is based on \autocite{hayes2014} (the same equation for both odd and even \(n\) values):

\[
A_n = -\dfrac{0.76213}{n} - \dfrac{0.86413}{n^2}.
\]

The second one is based on \autocite{williams2011}:

\[
A_n = -0.804168866 \cdot n^{-1.008922}.
\]

Both approaches produce almost identical results, so it does not actually matter which one to use.

\begin{figure}[ht!]

{\centering \includegraphics{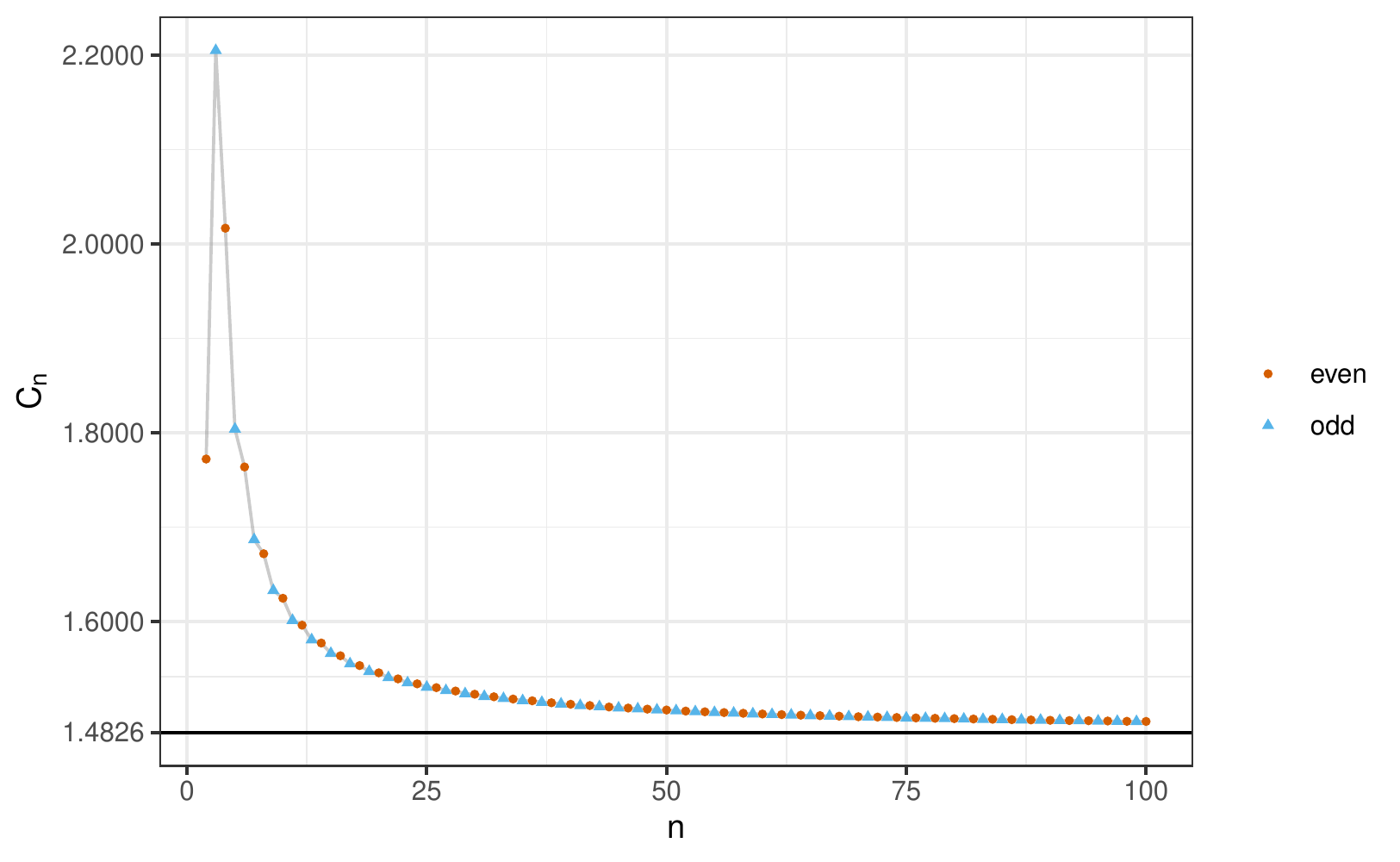} 

}

\caption{MAD bias-correction factors from the Park approach}\label{fig:parkPlot}
\end{figure}

For \(2 \leq n \leq 100\), they suggested to use predefined constants listed in Table~\ref{tab:park}
(based on Table A2 from \autocite{park2020}).
The corresponding plot is presented in Figure \ref{fig:parkPlot}.

\begin{table}[!h]

\caption{\label{tab:park}$C_n$ factors from the Park approach.}
\centering
\begin{tabular}[t]{r|r|r|r|r|r|r|r|r|r|r|r|r|r}
\hline
n & $C_n$ & n & $C_n$ & n & $C_n$ & n & $C_n$ & n & $C_n$ & n & $C_n$ & n & $C_n$\\
\hline
1 & - & 21 & 1.5407 & 41 & 1.5111 & 61 & 1.5016 & 81 & 1.4968 & 109 & 1.4931 & 249 & 1.4872\\
\hline
2 & 1.7722 & 22 & 1.5393 & 42 & 1.5110 & 62 & 1.5015 & 82 & 1.4968 & 110 & 1.4931 & 250 & 1.4872\\
\hline
3 & 2.2049 & 23 & 1.5352 & 43 & 1.5099 & 63 & 1.5010 & 83 & 1.4964 & 119 & 1.4923 & 299 & 1.4864\\
\hline
4 & 2.0167 & 24 & 1.5341 & 44 & 1.5095 & 64 & 1.5008 & 84 & 1.4965 & 120 & 1.4922 & 300 & 1.4864\\
\hline
5 & 1.8039 & 25 & 1.5305 & 45 & 1.5085 & 65 & 1.5003 & 85 & 1.4963 & 129 & 1.4914 & 349 & 1.4858\\
\hline
6 & 1.7638 & 26 & 1.5300 & 46 & 1.5084 & 66 & 1.5003 & 86 & 1.4961 & 130 & 1.4914 & 350 & 1.4858\\
\hline
7 & 1.6868 & 27 & 1.5269 & 47 & 1.5075 & 67 & 1.4999 & 87 & 1.4958 & 139 & 1.4908 & 399 & 1.4855\\
\hline
8 & 1.6718 & 28 & 1.5264 & 48 & 1.5072 & 68 & 1.4998 & 88 & 1.4958 & 140 & 1.4908 & 400 & 1.4855\\
\hline
9 & 1.6329 & 29 & 1.5236 & 49 & 1.5064 & 69 & 1.4993 & 89 & 1.4956 & 149 & 1.4902 & 449 & 1.4852\\
\hline
10 & 1.6247 & 30 & 1.5230 & 50 & 1.5063 & 70 & 1.4992 & 90 & 1.4954 & 150 & 1.4902 & 450 & 1.4852\\
\hline
11 & 1.6013 & 31 & 1.5207 & 51 & 1.5056 & 71 & 1.4989 & 91 & 1.4953 & 159 & 1.4897 & 499 & 1.4848\\
\hline
12 & 1.5962 & 32 & 1.5203 & 52 & 1.5052 & 72 & 1.4988 & 92 & 1.4951 & 160 & 1.4897 & 500 & 1.4848\\
\hline
13 & 1.5808 & 33 & 1.5185 & 53 & 1.5046 & 73 & 1.4985 & 93 & 1.4949 & 169 & 1.4893 & - & -\\
\hline
14 & 1.5773 & 34 & 1.5179 & 54 & 1.5044 & 74 & 1.4984 & 94 & 1.4950 & 170 & 1.4893 & - & -\\
\hline
15 & 1.5663 & 35 & 1.5163 & 55 & 1.5037 & 75 & 1.4979 & 95 & 1.4947 & 179 & 1.4890 & - & -\\
\hline
16 & 1.5638 & 36 & 1.5161 & 56 & 1.5036 & 76 & 1.4979 & 96 & 1.4947 & 180 & 1.4889 & - & -\\
\hline
17 & 1.5553 & 37 & 1.5144 & 57 & 1.5031 & 77 & 1.4975 & 97 & 1.4944 & 189 & 1.4887 & - & -\\
\hline
18 & 1.5534 & 38 & 1.5140 & 58 & 1.5029 & 78 & 1.4975 & 98 & 1.4943 & 190 & 1.4887 & - & -\\
\hline
19 & 1.5472 & 39 & 1.5127 & 59 & 1.5023 & 79 & 1.4972 & 99 & 1.4941 & 199 & 1.4883 & - & -\\
\hline
20 & 1.5457 & 40 & 1.5124 & 60 & 1.5021 & 80 & 1.4972 & 100 & 1.4942 & 200 & 1.4883 & - & -\\
\hline
\end{tabular}
\end{table}

\clearpage

\hypertarget{alternative-median-estimators}{%
\subsection{Alternative median estimators}\label{alternative-median-estimators}}

The described approach works quite well in practice for the sample median.
This estimator is the most robust median estimator (its breakdown point is 0.5),
but it does not have the best possible statistical efficiency
since it is based only on one or two order statistics.
Fortunately, there are other quantile estimators with better statistical efficiency.
One of the most popular alternatives which evaluate the median
as a weighted sum of all order statistics is the Harrell-Davis quantile estimator (see \autocite{harrell1982}).
Let \(Q(X, p)\) be an estimation of the \(p^\textrm{th}\) quantile of the random sample \(X\).
The Harrell-Davis quantile estimator \(Q_{\operatorname{HD}}(X, p)\) is defined as follows:

\[
Q_{\operatorname{HD}}(X, p) = \sum_{i=1}^{n} W_{\operatorname{HD},i} \cdot X_{(i)},\quad
W_{\operatorname{HD},i} = I_{i/n}(\alpha, \beta) - I_{(i-1)/n}(\alpha, \beta),
\]

where \(I_v(\alpha, \beta)\) is the regularized incomplete beta function,
\(\alpha = (n+1)p\), \(\;\beta = (n+1)(1-p)\),
\(X_{(i)}\) is the \(i^\textrm{th}\) order statistic of \(X\).

The Harrell-Davis quantile estimator is suggested in
\autocite{david2003}, \autocite{grissom2005}, \autocite{wilcox2016}, and \autocite{gibbons2020}
as an efficient alternative to the sample median.
In \autocite{yoshizawa1985} the Harrell-Davis median estimator is shown to be
asymptotically equivalent to the sample median.
While \(Q_{\operatorname{HD}}\) has great statistical efficiency,
it is not robust (its breakdown point is zero).
In practice, we still can use \(Q_{\operatorname{HD}}\) for medium-size outliers
without loss of accuracy because
the corresponding \(W_{\operatorname{HD},i}\) coefficients are quite small.
However, if a sample contains extreme outliers, \(Q_{\operatorname{HD}}\) can be corrupted.
Other examples of quantile estimators based on a weighted sum of all order statistics are
the Sfakianakis-Verginis quantile estimator (see \autocite{sfakianakis2008}) and
the Navruz-Özdemir quantile estimator (see \autocite{navruz2020}).
However, we continue considering only the Harrell-Davis quantile estimator
because it is the most popular option in this family.

In order to find an optimal trade-off between robustness and statistical efficiency,
we can consider the trimmed Harrell-Davis quantile estimator
based on the highest density interval of the given width that we denote by \(Q_{\operatorname{THD}}\)
(see \autocite{akinshin2022}).
In this modification of \(Q_{\operatorname{HD}}\),
we perform summation only within the highest density interval \([L;R]\) of \(\operatorname{Beta}(\alpha, \beta)\)
of size \(D\) (as a rule of thumb, we can use \(D = 1 / \sqrt{n}\) which gives us an estimator \(Q_{\operatorname{THD-SQRT}}\)).
It can be defined as follows:

\[
Q_{\operatorname{THD}}(X, p) = \sum_{i=1}^{n} W_{\operatorname{THD},i} \cdot X_{(i)}, \quad
W_{\operatorname{THD},i} = F_{\operatorname{THD}}(i / n) - F_{\operatorname{THD}}((i - 1) / n),
\]

\[
F_{\operatorname{THD}}(v) = \begin{cases}
0 & \textrm{for }\, v < L,\\
\big( I_v(\alpha, \beta) - I_L(\alpha, \beta) \big) /
\big( I_R(\alpha, \beta) \big) - I_L(\alpha, \beta) \big) \big)
  & \textrm{for }\, L \leq v \leq R,\\
1 & \textrm{for }\, R < v.
\end{cases}
\]

Quantile estimators \(Q_{\operatorname{HD}}\) and \(Q_{\operatorname{THD-SQRT}}\) can be also used as median estimators:
\(\operatorname{median}_{\operatorname{HD}}(X) = Q_{\operatorname{HD}}(X, 0.5)\),
\(\operatorname{median}_{\operatorname{THD-SQRT}}(X) = Q_{\operatorname{THD-SQRT}}(X, 0.5)\).
Let us denote the median absolute deviation based on \(Q_{\operatorname{HD}}\)
by \(\operatorname{MAD}_{\operatorname{HD}}\).
Similarly, we denote the median absolute deviation based on \(Q_{\operatorname{THD-SQRT}}\)
by \(\operatorname{MAD}_{\operatorname{THD-SQRT}}\).

In this paper, we conduct several simulation studies that evaluate approximated \(C_n\) values for
\(\operatorname{MAD}_{\operatorname{SM}}\),
\(\operatorname{MAD}_{\operatorname{THD}}\), and
\(\operatorname{MAD}_{\operatorname{THD-SQRT}}\).

\clearpage

\hypertarget{sec-simulations}{%
\section{Simulation study}\label{sec-simulations}}

In this section, we are going to perform several numerical simulations.
In Simulation~\protect\hyperlink{sim1}{1}, we get empirical values of the bias-correction factors \(C_n\)
for all considered \(\operatorname{MAD}\) estimators.
In Simulation~\protect\hyperlink{sim2}{2} and Simulation~\protect\hyperlink{sim3}{3}, we perform an analysis of
statistical efficiency and sensitivity to outliers
of the obtained unbiased estimators.

\hypertarget{sim1}{%
\subsection{Simulation 1: Evaluating bias-correction factors using the Monte-Carlo method}\label{sim1}}

Since \(C_n = 1/\mathbb{E}[\operatorname{MAD}_0(X)]\), this value can be obtained
by estimating the expected value of \(\operatorname{MAD}_0(X)\) using the Monte-Carlo method.
We do it according to the following scheme:

\begin{algorithm}[H]
\ForEach{$\textit{median}_* \in \{ \operatorname{median}_{\operatorname{SM}},\, \operatorname{median}_{\operatorname{HD}},\, \operatorname{median}_{\operatorname{THD-SQRT}}\}$}{
  \ForEach{$n \in \{ 2..100, \ldots, 3000 \}$}{
  $\textit{repetitions} \gets \textbf{when} \,\{ n \leq 10 \to 10^9;\; n \leq 100 \to 5\cdot10^8;\; \textbf{else} \to 2\cdot10^8 \}$\\
    \For{$i \gets 1..\textit{repetitions}$}{
         $x \gets \textrm{GenerateRandomSample}(\textrm{Distribution} = \mathcal{N}(0, 1^2),\, \textrm{SampleSize} = n)$\\
         $m_i \gets \textit{median}_*(|x-\textit{median}_*(x)|)$ 
    }
   $M_n \gets \sum m_i / \textit{repetitions}$\\
   $C_n \gets 1 / M_n$
  }
}
\end{algorithm}

\begin{figure}[ht!]

{\centering \includegraphics{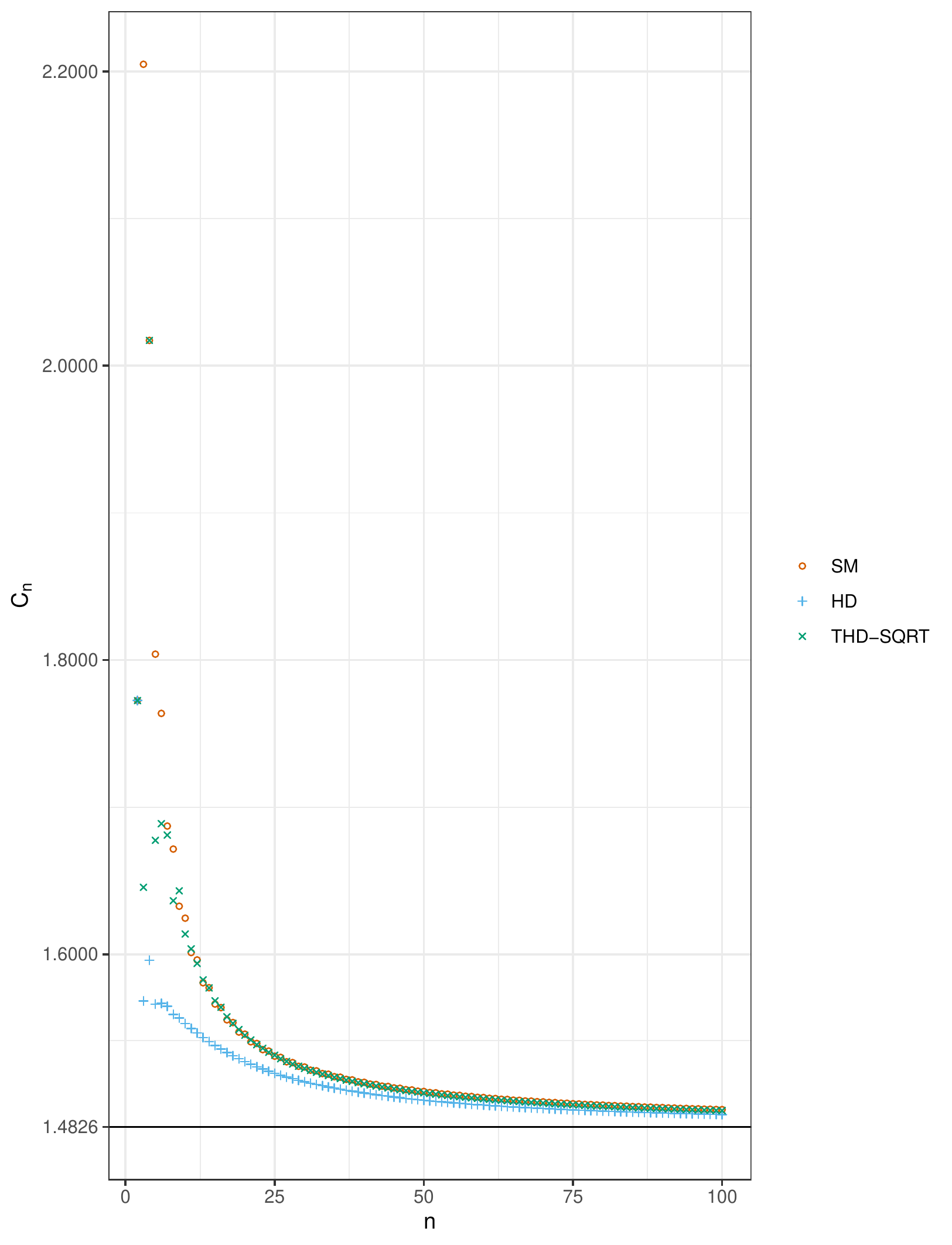} 

}

\caption{Bias-correction factors using different median estimators.}\label{fig:factorPlot}
\end{figure}

The estimated \(C_n\) values for
\(\operatorname{MAD}_{\operatorname{SM}}\), \(\operatorname{MAD}_{\operatorname{HD}}\), \(\operatorname{MAD}_{\operatorname{THD-SQRT}}\)
are presented in Tables~\ref{tab:sm}, \ref{tab:hd}, and \ref{tab:thd-sqrt} respectively.
A visualization for \(2 \leq n \leq 100\) is shown in Figure \ref{fig:factorPlot}.
The simulation for \(\operatorname{MAD}_{\operatorname{SM}}\) replicates the study from \autocite{park2020}
with a higher number of samples (they used \(10^7\) random samples).
The results of two studies (Tables~\ref{tab:park} and \ref{tab:sm}) are quite close to each other
(the maximum observed absolute difference is \(\approx 0.00065\)).

\begin{table}[!h]

\caption{\label{tab:sm}$C_n$ factors for $\operatorname{MAD}_{\operatorname{SM}}$.}
\centering
\begin{tabular}[t]{r|r|r|r|r|r|r|r|r|r|r|r|r|r}
\hline
n & $C_n$ & n & $C_n$ & n & $C_n$ & n & $C_n$ & n & $C_n$ & n & $C_n$ & n & $C_n$\\
\hline
1 & - & 21 & 1.5405 & 41 & 1.5117 & 61 & 1.5021 & 81 & 1.4972 & 109 & 1.4933 & 249 & 1.4872\\
\hline
2 & 1.7725 & 22 & 1.5393 & 42 & 1.5115 & 62 & 1.5019 & 82 & 1.4971 & 110 & 1.4933 & 250 & 1.4872\\
\hline
3 & 2.2049 & 23 & 1.5352 & 43 & 1.5103 & 63 & 1.5014 & 83 & 1.4968 & 119 & 1.4924 & 299 & 1.4864\\
\hline
4 & 2.0172 & 24 & 1.5342 & 44 & 1.5101 & 64 & 1.5013 & 84 & 1.4967 & 120 & 1.4924 & 300 & 1.4864\\
\hline
5 & 1.8040 & 25 & 1.5307 & 45 & 1.5091 & 65 & 1.5008 & 85 & 1.4965 & 129 & 1.4916 & 349 & 1.4859\\
\hline
6 & 1.7637 & 26 & 1.5299 & 46 & 1.5089 & 66 & 1.5007 & 86 & 1.4964 & 130 & 1.4916 & 350 & 1.4859\\
\hline
7 & 1.6871 & 27 & 1.5269 & 47 & 1.5080 & 67 & 1.5003 & 87 & 1.4961 & 139 & 1.4910 & 399 & 1.4855\\
\hline
8 & 1.6715 & 28 & 1.5263 & 48 & 1.5078 & 68 & 1.5002 & 88 & 1.4961 & 140 & 1.4910 & 400 & 1.4855\\
\hline
9 & 1.6326 & 29 & 1.5238 & 49 & 1.5069 & 69 & 1.4998 & 89 & 1.4958 & 149 & 1.4904 & 449 & 1.4852\\
\hline
10 & 1.6245 & 30 & 1.5233 & 50 & 1.5067 & 70 & 1.4997 & 90 & 1.4958 & 150 & 1.4904 & 450 & 1.4851\\
\hline
11 & 1.6011 & 31 & 1.5212 & 51 & 1.5060 & 71 & 1.4993 & 91 & 1.4955 & 159 & 1.4899 & 499 & 1.4849\\
\hline
12 & 1.5961 & 32 & 1.5207 & 52 & 1.5058 & 72 & 1.4992 & 92 & 1.4955 & 160 & 1.4899 & 500 & 1.4849\\
\hline
13 & 1.5806 & 33 & 1.5189 & 53 & 1.5051 & 73 & 1.4988 & 93 & 1.4952 & 169 & 1.4895 & 600 & 1.4845\\
\hline
14 & 1.5772 & 34 & 1.5184 & 54 & 1.5049 & 74 & 1.4987 & 94 & 1.4952 & 170 & 1.4895 & 700 & 1.4842\\
\hline
15 & 1.5661 & 35 & 1.5168 & 55 & 1.5042 & 75 & 1.4984 & 95 & 1.4950 & 179 & 1.4891 & 800 & 1.4840\\
\hline
16 & 1.5637 & 36 & 1.5164 & 56 & 1.5041 & 76 & 1.4983 & 96 & 1.4949 & 180 & 1.4891 & 900 & 1.4839\\
\hline
17 & 1.5554 & 37 & 1.5149 & 57 & 1.5035 & 77 & 1.4979 & 97 & 1.4947 & 189 & 1.4887 & 1000 & 1.4837\\
\hline
18 & 1.5536 & 38 & 1.5146 & 58 & 1.5033 & 78 & 1.4978 & 98 & 1.4947 & 190 & 1.4887 & 1500 & 1.4834\\
\hline
19 & 1.5471 & 39 & 1.5132 & 59 & 1.5027 & 79 & 1.4975 & 99 & 1.4945 & 199 & 1.4884 & 2000 & 1.4832\\
\hline
20 & 1.5457 & 40 & 1.5129 & 60 & 1.5026 & 80 & 1.4975 & 100 & 1.4944 & 200 & 1.4884 & 3000 & 1.4830\\
\hline
\end{tabular}
\end{table}

\clearpage

\begin{table}[!h]

\caption{\label{tab:hd}$C_n$ factors for $\operatorname{MAD}_{\operatorname{HD}}$.}
\centering
\begin{tabular}[t]{r|r|r|r|r|r|r|r|r|r|r|r|r|r}
\hline
n & $C_n$ & n & $C_n$ & n & $C_n$ & n & $C_n$ & n & $C_n$ & n & $C_n$ & n & $C_n$\\
\hline
1 & - & 21 & 1.5252 & 41 & 1.5050 & 61 & 1.4972 & 81 & 1.4933 & 109 & 1.4902 & 249 & 1.4857\\
\hline
2 & 1.7725 & 22 & 1.5235 & 42 & 1.5045 & 62 & 1.4969 & 82 & 1.4931 & 110 & 1.4902 & 250 & 1.4857\\
\hline
3 & 1.5682 & 23 & 1.5220 & 43 & 1.5039 & 63 & 1.4967 & 83 & 1.4930 & 119 & 1.4896 & 299 & 1.4852\\
\hline
4 & 1.5959 & 24 & 1.5204 & 44 & 1.5034 & 64 & 1.4964 & 84 & 1.4928 & 120 & 1.4895 & 300 & 1.4852\\
\hline
5 & 1.5661 & 25 & 1.5191 & 45 & 1.5029 & 65 & 1.4962 & 85 & 1.4927 & 129 & 1.4890 & 349 & 1.4848\\
\hline
6 & 1.5666 & 26 & 1.5177 & 46 & 1.5025 & 66 & 1.4960 & 86 & 1.4926 & 130 & 1.4889 & 350 & 1.4848\\
\hline
7 & 1.5646 & 27 & 1.5164 & 47 & 1.5020 & 67 & 1.4957 & 87 & 1.4924 & 139 & 1.4884 & 399 & 1.4845\\
\hline
8 & 1.5591 & 28 & 1.5154 & 48 & 1.5016 & 68 & 1.4955 & 88 & 1.4923 & 140 & 1.4884 & 400 & 1.4845\\
\hline
9 & 1.5567 & 29 & 1.5143 & 49 & 1.5011 & 69 & 1.4953 & 89 & 1.4922 & 149 & 1.4880 & 449 & 1.4843\\
\hline
10 & 1.5529 & 30 & 1.5133 & 50 & 1.5008 & 70 & 1.4951 & 90 & 1.4921 & 150 & 1.4880 & 450 & 1.4843\\
\hline
11 & 1.5496 & 31 & 1.5123 & 51 & 1.5004 & 71 & 1.4950 & 91 & 1.4920 & 159 & 1.4877 & 499 & 1.4841\\
\hline
12 & 1.5465 & 32 & 1.5114 & 52 & 1.5000 & 72 & 1.4947 & 92 & 1.4918 & 160 & 1.4876 & 500 & 1.4841\\
\hline
13 & 1.5434 & 33 & 1.5106 & 53 & 1.4997 & 73 & 1.4946 & 93 & 1.4917 & 169 & 1.4873 & 600 & 1.4838\\
\hline
14 & 1.5406 & 34 & 1.5098 & 54 & 1.4993 & 74 & 1.4944 & 94 & 1.4916 & 170 & 1.4873 & 700 & 1.4836\\
\hline
15 & 1.5380 & 35 & 1.5090 & 55 & 1.4990 & 75 & 1.4942 & 95 & 1.4915 & 179 & 1.4871 & 800 & 1.4835\\
\hline
16 & 1.5355 & 36 & 1.5083 & 56 & 1.4986 & 76 & 1.4940 & 96 & 1.4914 & 180 & 1.4870 & 900 & 1.4834\\
\hline
17 & 1.5332 & 37 & 1.5076 & 57 & 1.4983 & 77 & 1.4939 & 97 & 1.4913 & 189 & 1.4868 & 1000 & 1.4833\\
\hline
18 & 1.5310 & 38 & 1.5069 & 58 & 1.4980 & 78 & 1.4937 & 98 & 1.4912 & 190 & 1.4868 & 1500 & 1.4831\\
\hline
19 & 1.5289 & 39 & 1.5062 & 59 & 1.4977 & 79 & 1.4936 & 99 & 1.4911 & 199 & 1.4866 & 2000 & 1.4830\\
\hline
20 & 1.5270 & 40 & 1.5056 & 60 & 1.4975 & 80 & 1.4934 & 100 & 1.4910 & 200 & 1.4866 & 3000 & 1.4828\\
\hline
\end{tabular}
\end{table}

\begin{table}[!h]

\caption{\label{tab:thd-sqrt}$C_n$ factors for $\operatorname{MAD}_{\operatorname{THD-SQRT}}$.}
\centering
\begin{tabular}[t]{r|r|r|r|r|r|r|r|r|r|r|r|r|r}
\hline
n & $C_n$ & n & $C_n$ & n & $C_n$ & n & $C_n$ & n & $C_n$ & n & $C_n$ & n & $C_n$\\
\hline
1 & - & 21 & 1.5417 & 41 & 1.5111 & 61 & 1.5013 & 81 & 1.4965 & 109 & 1.4927 & 249 & 1.4869\\
\hline
2 & 1.7725 & 22 & 1.5385 & 42 & 1.5104 & 62 & 1.5010 & 82 & 1.4963 & 110 & 1.4926 & 250 & 1.4869\\
\hline
3 & 1.6455 & 23 & 1.5361 & 43 & 1.5097 & 63 & 1.5007 & 83 & 1.4961 & 119 & 1.4919 & 299 & 1.4861\\
\hline
4 & 2.0172 & 24 & 1.5333 & 44 & 1.5091 & 64 & 1.5004 & 84 & 1.4959 & 120 & 1.4918 & 300 & 1.4861\\
\hline
5 & 1.6774 & 25 & 1.5313 & 45 & 1.5085 & 65 & 1.5001 & 85 & 1.4958 & 129 & 1.4911 & 349 & 1.4856\\
\hline
6 & 1.6887 & 26 & 1.5290 & 46 & 1.5078 & 66 & 1.4998 & 86 & 1.4956 & 130 & 1.4910 & 350 & 1.4856\\
\hline
7 & 1.6810 & 27 & 1.5272 & 47 & 1.5073 & 67 & 1.4995 & 87 & 1.4955 & 139 & 1.4904 & 399 & 1.4852\\
\hline
8 & 1.6363 & 28 & 1.5254 & 48 & 1.5067 & 68 & 1.4993 & 88 & 1.4953 & 140 & 1.4904 & 400 & 1.4852\\
\hline
9 & 1.6431 & 29 & 1.5238 & 49 & 1.5063 & 69 & 1.4990 & 89 & 1.4952 & 149 & 1.4899 & 449 & 1.4849\\
\hline
10 & 1.6137 & 30 & 1.5224 & 50 & 1.5057 & 70 & 1.4988 & 90 & 1.4950 & 150 & 1.4898 & 450 & 1.4849\\
\hline
11 & 1.6036 & 31 & 1.5210 & 51 & 1.5053 & 71 & 1.4986 & 91 & 1.4949 & 159 & 1.4894 & 499 & 1.4847\\
\hline
12 & 1.5938 & 32 & 1.5198 & 52 & 1.5048 & 72 & 1.4983 & 92 & 1.4947 & 160 & 1.4894 & 500 & 1.4847\\
\hline
13 & 1.5826 & 33 & 1.5185 & 53 & 1.5044 & 73 & 1.4981 & 93 & 1.4946 & 169 & 1.4890 & 600 & 1.4843\\
\hline
14 & 1.5771 & 34 & 1.5175 & 54 & 1.5039 & 74 & 1.4979 & 94 & 1.4944 & 170 & 1.4890 & 700 & 1.4841\\
\hline
15 & 1.5683 & 35 & 1.5163 & 55 & 1.5035 & 75 & 1.4977 & 95 & 1.4943 & 179 & 1.4886 & 800 & 1.4839\\
\hline
16 & 1.5639 & 36 & 1.5155 & 56 & 1.5031 & 76 & 1.4974 & 96 & 1.4942 & 180 & 1.4886 & 900 & 1.4837\\
\hline
17 & 1.5574 & 37 & 1.5144 & 57 & 1.5027 & 77 & 1.4972 & 97 & 1.4940 & 189 & 1.4883 & 1000 & 1.4836\\
\hline
18 & 1.5530 & 38 & 1.5136 & 58 & 1.5024 & 78 & 1.4970 & 98 & 1.4940 & 190 & 1.4883 & 1500 & 1.4833\\
\hline
19 & 1.5488 & 39 & 1.5127 & 59 & 1.5020 & 79 & 1.4969 & 99 & 1.4938 & 199 & 1.4880 & 2000 & 1.4831\\
\hline
20 & 1.5449 & 40 & 1.5119 & 60 & 1.5017 & 80 & 1.4966 & 100 & 1.4937 & 200 & 1.4880 & 3000 & 1.4829\\
\hline
\end{tabular}
\end{table}

\clearpage

\clearpage

\hypertarget{sim2}{%
\subsection{Simulation 2: Statistical efficiency of the median absolute deviation}\label{sim2}}

In this simulation, we estimate the relative efficiency \(e\) of
\(\operatorname{MAD}_{\operatorname{HD}}\) and \(\operatorname{MAD}_{\operatorname{THD-SQRT}}\)
against \(\operatorname{MAD}_{\operatorname{SM}}\) (the baseline).
It can be calculated as the ratio of the estimator mean squared errors (\(\operatorname{MSE}\)) (see \autocite{dekking2005}).
Since all the estimators are unbiased under normality,
\(\operatorname{MSE}(\operatorname{MAD}_*) = \mathbb{V}[\operatorname{MAD}_*]\).
Thus, we have:

\[
e(\operatorname{MAD}_*) =
  \dfrac{\operatorname{MSE}(\operatorname{MAD}_{\operatorname{SM}})}{\operatorname{MSE}(\operatorname{MAD}_*)} =
  \dfrac{\mathbb{V}[\operatorname{MAD}_{\operatorname{SM}}(X)]}{\mathbb{V}[\operatorname{MAD}_*(X)]},
\]

where \(\mathbb{V}\) is the variance of \(\operatorname{MAD}_n\) for the given sample size \(n\),
\(\operatorname{MAD}_*\) is a placeholder for \(\operatorname{MAD}_{\operatorname{HD}}\) and \(\operatorname{MAD}_{\operatorname{THD-SQRT}}\).
We conduct this simulation according to the following scheme:

\begin{algorithm}[H]
\ForEach{$n \in \{ 2, 3, 4, 5, 6, 7, 10, 50, 100, 500, 1000 \}$}{
  \For{$i \gets 1..10\,000$}{
    $x \gets \textrm{GenerateRandomSample}(\textrm{Distribution} = \mathcal{N}(0, 1^2),\, \textrm{SampleSize} = n)$\\
    $m_{\operatorname{SM},i} = \operatorname{MAD}_{\operatorname{SM},n}(x)$\\
    $m_{\operatorname{HD},i} = \operatorname{MAD}_{\operatorname{HD},n}(x)$\\
    $m_{\operatorname{THD-SQRT},i} = \operatorname{MAD}_{\operatorname{THD-SQRT},n}(x)$\\
  }
  $e(\operatorname{MAD}_{\operatorname{HD},n}) = \mathbb{V}(m_{\operatorname{SM},\{i\}}) / \mathbb{V}(m_{\operatorname{HD},\{i\}})$\\
  $e(\operatorname{MAD}_{\operatorname{THD-SQRT},n}) = \mathbb{V}(m_{\operatorname{SM},\{i\}}) / \mathbb{V}(m_{\operatorname{THD-SQRT},\{i\}})$\\
}
\end{algorithm}

The evaluated values of the
\(e(\operatorname{MAD}_{\operatorname{HD}})\) and \(e(\operatorname{MAD}_{\operatorname{THD-SQRT}})\)
are presented in Table~\ref{tab:effSummary}.

\begin{table}[!h]

\caption{\label{tab:effSummary}Relative statistical efficiency of the median absolute deviation.}
\centering
\begin{tabular}[t]{r|r|r}
\hline
n & HD & THD-SQRT\\
\hline
2 & 1.000 & 1.000\\
\hline
3 & 2.473 & 2.331\\
\hline
4 & 1.618 & 1.000\\
\hline
5 & 1.854 & 1.468\\
\hline
6 & 1.473 & 1.180\\
\hline
7 & 1.688 & 1.326\\
\hline
8 & 1.379 & 1.141\\
\hline
9 & 1.527 & 1.227\\
\hline
10 & 1.342 & 1.129\\
\hline
50 & 1.156 & 1.075\\
\hline
100 & 1.110 & 1.054\\
\hline
500 & 1.047 & 1.025\\
\hline
1000 & 1.035 & 1.018\\
\hline
\end{tabular}
\end{table}

Based on the obtained measurements, we can do the following observations
about the efficiency of the considered \(\operatorname{MAD}\) estimators
\emph{under the normal distribution}:

\begin{itemize}
\tightlist
\item
  Both \(\operatorname{MAD}_{\operatorname{HD}}\) and \(\operatorname{MAD}_{\operatorname{THD-SQRT}}\)
  are more efficient than \(\operatorname{MAD}_{\operatorname{SM}}\).
\item
  \(\operatorname{MAD}_{\operatorname{HD}}\) is more efficient than \(\operatorname{MAD}_{\operatorname{THD-SQRT}}\).
\item
  The impact of using \(\operatorname{MAD}_{\operatorname{HD}}\) and \(\operatorname{MAD}_{\operatorname{THD-SQRT}}\)
  instead of \(\operatorname{MAD}_{\operatorname{SM}}\)
  is most noticeable for small samples
  (except \(n=2\) for \(\operatorname{MAD}_{\operatorname{HD}}\) and
  \(n \in \{ 2,4 \}\) for \(\operatorname{MAD}_{\operatorname{THD-SQRT}}\)).
  The most impressive boost of efficiency can be observed for \(n=3\):
  \(+147.3\%\) for \(\operatorname{MAD}_{\operatorname{HD}}\) and
  \(+133.1\%\) for \(\operatorname{MAD}_{\operatorname{THD-SQRT}}\).
\item
  For large samples, \(\operatorname{MAD}_{\operatorname{HD}}\) and \(\operatorname{MAD}_{\operatorname{THD-SQRT}}\)
  are still more efficient than \(\operatorname{MAD}_{\operatorname{SM}}\),
  but the difference is not so noticeable.
  For example, for \(n=1000\),
  \(\operatorname{MAD}_{\operatorname{HD}}\) gives \(+3.5\%\) and
  \(\operatorname{MAD}_{\operatorname{THD-SQRT}}\) gives \(+1.8\%\) to statistical efficiency.
\end{itemize}

\clearpage

\hypertarget{sim3}{%
\subsection{Simulation 3: Sensitivity to outliers of the median absolute deviation}\label{sim3}}

There are various metrics that describe robustness
(e.g., the breakdown point, the influence function, and the sensitivity curve).
While these metrics provide important theoretical properties,
they do not present a clear visual illustration
of the actual impact of outliers on estimations.
Another approach to getting an idea of the sensitivity of different estimators to outliers
is exploring the statistical dispersion of obtained estimations on light-tailed and heavy-tailed distributions.
Let us conduct a simulation according to the following scheme:

\begin{algorithm}[H]
\ForEach{$d \in \mathcal{D}$}{
\ForEach{$n \in \{ 2, 3, 4, 5, 6, 7, 8, 9, 10, 50, 100, 500, 1000 \}$}{
  \For{$i \gets 1..1\,000$}{
     $x \gets \textrm{GenerateRandomSample}(\textrm{Distribution} = d,\, \textrm{SampleSize} = n)$\\
     \ForEach{$\textit{estimator} \in \{ \operatorname{SM}, \operatorname{HD}, \operatorname{THD-SQRT} \} $}{
        $m_{\textit{estimator},i} = \operatorname{MAD}_{\textit{estimator},n}(x)$\\
     }
  }
  \ForEach{$\textit{aggregator} \in \{ \operatorname{SD}, \operatorname{IQR}, \operatorname{MAD}_{\operatorname{SM}} \}$}{
    $\operatorname{Result}(d, n, \textit{estimator}, \textit{aggregator}) = \textit{aggregator}(m_{\textit{estimator},\{i\}})$
  }
}
}   
\end{algorithm}

\begin{figure}[ht!]

{\centering \includegraphics{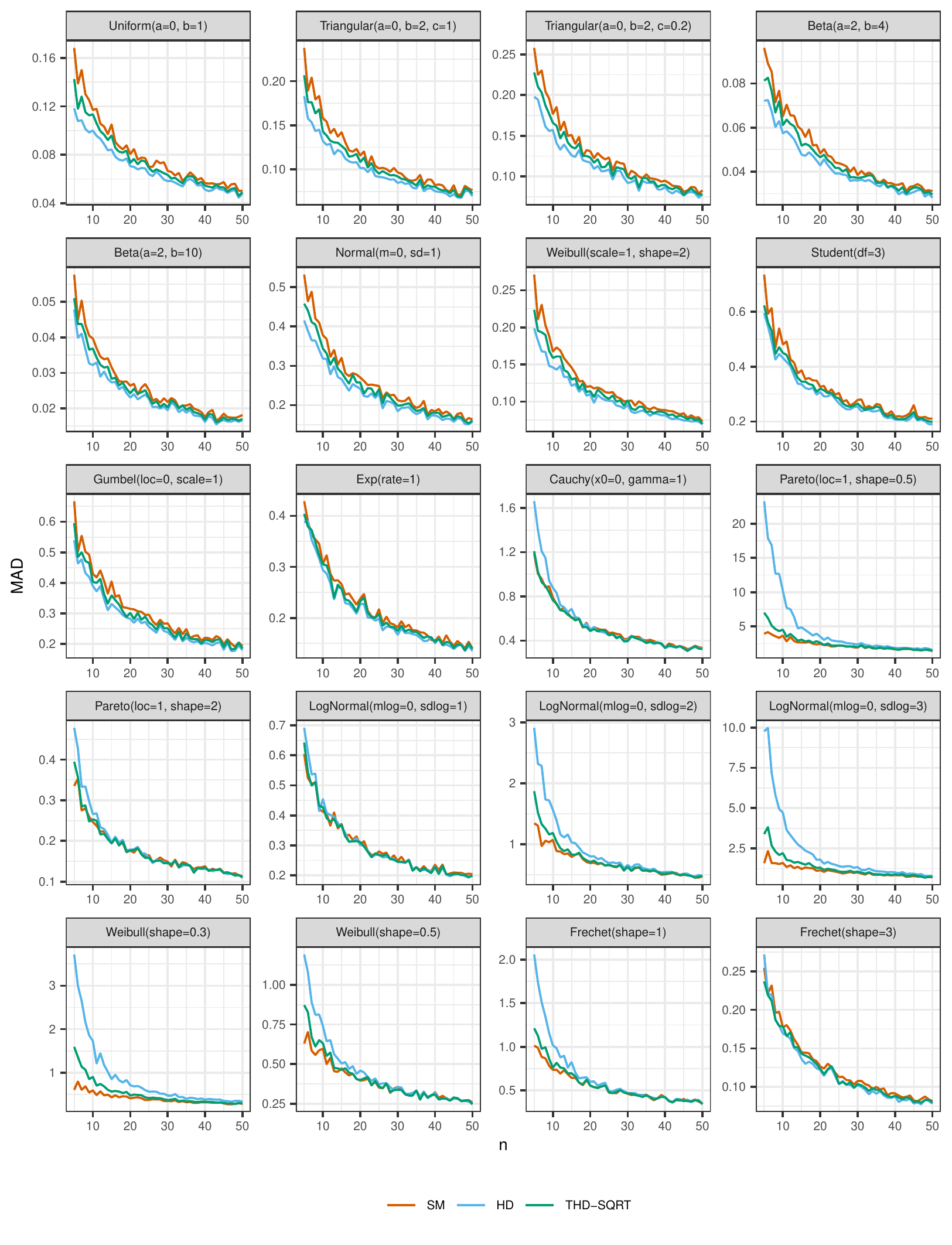} 

}

\caption{Statistical dispersion of MAD estimations on various distributions.}\label{fig:robust}
\end{figure}

In this simulation, we enumerate a set \(\mathcal{D}\) of distributions listed in Table \ref{tab:ds}
(this set includes symmetric and skewed, light-tailed and heavy-tailed distributions).
We describe the statistical dispersion of each set of \(\operatorname{MAD}_n\) estimations in three different ways:
\(\operatorname{SD}\) (the classic standard deviation),
\(\operatorname{IQR}\) (interquartile range based on Hyndman-Fan Type 7 quantile estimator),
\(\operatorname{MAD}_{\operatorname{SM}}\).
The aggregated results for \(n \in \{ 5, 6, 10, 50 \}\) are listed
in Tables~\ref{tab:robust5}, \ref{tab:robust6},\ref{tab:robust10}, \ref{tab:robust50} respectively.
The \(\operatorname{MAD}_{\operatorname{SM}}\)-aggregated results for all values of \(n\) are presented in Figure~\ref{fig:robust}.
Based on the obtained measurements, we can make the following observations:

\begin{itemize}
\tightlist
\item
  For the light-tailed distributions,
  \(\operatorname{MAD}_{\operatorname{HD}}\) has the best robustness,
  \(\operatorname{MAD}_{\operatorname{SM}}\) has the worst robustness.
\item
  For the heavy-tailed distributions, the opposite is true:
  \(\operatorname{MAD}_{\operatorname{HD}}\) has the worst robustness,
  \(\operatorname{MAD}_{\operatorname{SM}}\) has the best robustness.
  On small samples,
  \(\operatorname{MAD}_{\operatorname{HD}}\) could be much worse than \(\operatorname{MAD}_{\operatorname{SM}}\)
  while \(\operatorname{MAD}_{\operatorname{THD-SQRT}}\) is just a little bit worse.
\item
  For \(n \geq 50\), the difference between all considered estimators is negligible.
\end{itemize}

\begin{longtable}[]{@{}
  >{\raggedright\arraybackslash}p{(\columnwidth - 6\tabcolsep) * \real{0.3718}}
  >{\raggedright\arraybackslash}p{(\columnwidth - 6\tabcolsep) * \real{0.2692}}
  >{\raggedright\arraybackslash}p{(\columnwidth - 6\tabcolsep) * \real{0.1795}}
  >{\raggedright\arraybackslash}p{(\columnwidth - 6\tabcolsep) * \real{0.1795}}@{}}
\caption{\label{tab:ds} Distributions for Simulation~\protect\hyperlink{sim3}{3}.}\tabularnewline
\toprule
\begin{minipage}[b]{\linewidth}\raggedright
Distribution
\end{minipage} & \begin{minipage}[b]{\linewidth}\raggedright
Support
\end{minipage} & \begin{minipage}[b]{\linewidth}\raggedright
Skewness
\end{minipage} & \begin{minipage}[b]{\linewidth}\raggedright
Tailness
\end{minipage} \\
\midrule
\endfirsthead
\toprule
\begin{minipage}[b]{\linewidth}\raggedright
Distribution
\end{minipage} & \begin{minipage}[b]{\linewidth}\raggedright
Support
\end{minipage} & \begin{minipage}[b]{\linewidth}\raggedright
Skewness
\end{minipage} & \begin{minipage}[b]{\linewidth}\raggedright
Tailness
\end{minipage} \\
\midrule
\endhead
Uniform(a=0, b=1) & \([0;1]\) & Symmetric & Light-tailed \\
Triangular(a=0, b=2, c=1) & \([0;2]\) & Symmetric & Light-tailed \\
Triangular(a=0, b=2, c=0.2) & \([0;2]\) & Right-skewed & Light-tailed \\
Beta(a=2, b=4) & \([0;1]\) & Right-skewed & Light-tailed \\
Beta(a=2, b=10) & \([0;1]\) & Right-skewed & Light-tailed \\
Normal(m=0, sd=1) & \((-\infty;+\infty)\) & Symmetric & Light-tailed \\
Weibull(scale=1, shape=2) & \([0;+\infty)\) & Right-skewed & Light-tailed \\
Student(df=3) & \((-\infty;+\infty)\) & Symmetric & Light-tailed \\
Gumbel(loc=0, scale=1) & \((-\infty;+\infty)\) & Right-skewed & Light-tailed \\
Exp(rate=1) & \([0;+\infty)\) & Right-skewed & Light-tailed \\
Cauchy(x0=0, gamma=1) & \((-\infty;+\infty)\) & Symmetric & Heavy-tailed \\
Pareto(loc=1, shape=0.5) & \([1;+\infty)\) & Right-skewed & Heavy-tailed \\
Pareto(loc=1, shape=2) & \([1;+\infty)\) & Right-skewed & Heavy-tailed \\
LogNormal(mlog=0, sdlog=1) & \((0;+\infty)\) & Right-skewed & Heavy-tailed \\
LogNormal(mlog=0, sdlog=2) & \((0;+\infty)\) & Right-skewed & Heavy-tailed \\
LogNormal(mlog=0, sdlog=3) & \((0;+\infty)\) & Right-skewed & Heavy-tailed \\
Weibull(shape=0.3) & \([0;+\infty)\) & Right-skewed & Heavy-tailed \\
Weibull(shape=0.5) & \([0;+\infty)\) & Right-skewed & Heavy-tailed \\
Frechet(shape=1) & \((0;+\infty)\) & Right-skewed & Heavy-tailed \\
Frechet(shape=3) & \((0;+\infty)\) & Right-skewed & Heavy-tailed \\
\bottomrule
\end{longtable}

\clearpage

\begin{table}[!h]

\caption{\label{tab:robust5}Properties of MAD estimations for n=5.}
\centering
\begin{tabular}[t]{l|r|r|r|r|r|r|r|r|r}
\hline
\multicolumn{1}{c|}{ } & \multicolumn{3}{c|}{SD} & \multicolumn{3}{c|}{IQR} & \multicolumn{3}{c}{MAD} \\
\cline{2-4} \cline{5-7} \cline{8-10}
Distribution & SM & HD & THD & SM & HD & THD & SM & HD & THD\\
\hline
Uniform(a=0, b=1) & 0.16 & 0.12 & 0.14 & 0.23 & 0.16 & 0.19 & 0.17 & 0.12 & 0.14\\
\hline
Triangular(a=0, b=2, c=1) & 0.24 & 0.17 & 0.19 & 0.33 & 0.25 & 0.28 & 0.24 & 0.18 & 0.21\\
\hline
Triangular(a=0, b=2, c=0.2) & 0.25 & 0.20 & 0.22 & 0.36 & 0.27 & 0.31 & 0.26 & 0.20 & 0.23\\
\hline
Beta(a=2, b=4) & 0.10 & 0.07 & 0.08 & 0.13 & 0.10 & 0.11 & 0.10 & 0.07 & 0.08\\
\hline
Beta(a=2, b=10) & 0.06 & 0.05 & 0.05 & 0.08 & 0.07 & 0.07 & 0.06 & 0.05 & 0.05\\
\hline
Normal(m=0, sd=1) & 0.59 & 0.43 & 0.48 & 0.72 & 0.57 & 0.65 & 0.53 & 0.42 & 0.46\\
\hline
Weibull(scale=1, shape=2) & 0.26 & 0.20 & 0.22 & 0.38 & 0.27 & 0.30 & 0.27 & 0.20 & 0.22\\
\hline
Student(df=3) & 0.79 & 0.75 & 0.74 & 1.00 & 0.83 & 0.86 & 0.74 & 0.60 & 0.62\\
\hline
Gumbel(loc=0, scale=1) & 0.77 & 0.58 & 0.63 & 0.93 & 0.76 & 0.83 & 0.67 & 0.54 & 0.60\\
\hline
Exp(rate=1) & 0.51 & 0.44 & 0.46 & 0.63 & 0.54 & 0.56 & 0.43 & 0.39 & 0.40\\
\hline
Cauchy(x0=0, gamma=1) & 1.90 & 10.92 & 2.41 & 1.70 & 3.03 & 1.84 & 1.18 & 1.66 & 1.21\\
\hline
Pareto(loc=1, shape=0.5) & 26.80 & 1645746.00 & 154.80 & 7.68 & 68.61 & 15.27 & 3.94 & 23.28 & 7.01\\
\hline
Pareto(loc=1, shape=2) & 0.56 & 1.03 & 0.63 & 0.53 & 0.69 & 0.57 & 0.34 & 0.48 & 0.39\\
\hline
LogNormal(mlog=0, sdlog=1) & 0.85 & 0.99 & 0.91 & 0.86 & 1.00 & 0.91 & 0.60 & 0.69 & 0.64\\
\hline
LogNormal(mlog=0, sdlog=2) & 3.32 & 18.33 & 4.39 & 2.17 & 5.57 & 3.04 & 1.34 & 2.91 & 1.87\\
\hline
LogNormal(mlog=0, sdlog=3) & 14.53 & 833.22 & 31.64 & 3.83 & 24.28 & 7.20 & 1.59 & 9.76 & 3.39\\
\hline
Weibull(shape=0.3) & 3.81 & 16.26 & 7.23 & 1.42 & 7.50 & 3.36 & 0.60 & 3.72 & 1.59\\
\hline
Weibull(shape=0.5) & 1.45 & 2.08 & 1.73 & 1.14 & 1.87 & 1.38 & 0.63 & 1.19 & 0.87\\
\hline
Frechet(shape=1) & 4.52 & 52.16 & 9.99 & 1.65 & 3.62 & 1.87 & 1.01 & 2.06 & 1.21\\
\hline
Frechet(shape=3) & 0.33 & 0.37 & 0.33 & 0.36 & 0.38 & 0.34 & 0.25 & 0.27 & 0.24\\
\hline
\end{tabular}
\end{table}

\begin{table}[!h]

\caption{\label{tab:robust6}Properties of MAD estimations for n=6.}
\centering
\begin{tabular}[t]{l|r|r|r|r|r|r|r|r|r}
\hline
\multicolumn{1}{c|}{ } & \multicolumn{3}{c|}{SD} & \multicolumn{3}{c|}{IQR} & \multicolumn{3}{c}{MAD} \\
\cline{2-4} \cline{5-7} \cline{8-10}
Distribution & SM & HD & THD & SM & HD & THD & SM & HD & THD\\
\hline
Uniform(a=0, b=1) & 0.14 & 0.11 & 0.12 & 0.19 & 0.14 & 0.16 & 0.14 & 0.11 & 0.12\\
\hline
Triangular(a=0, b=2, c=1) & 0.19 & 0.16 & 0.18 & 0.25 & 0.21 & 0.24 & 0.19 & 0.16 & 0.18\\
\hline
Triangular(a=0, b=2, c=0.2) & 0.22 & 0.18 & 0.20 & 0.30 & 0.27 & 0.29 & 0.22 & 0.19 & 0.21\\
\hline
Beta(a=2, b=4) & 0.09 & 0.07 & 0.08 & 0.12 & 0.10 & 0.11 & 0.09 & 0.07 & 0.08\\
\hline
Beta(a=2, b=10) & 0.05 & 0.04 & 0.04 & 0.06 & 0.05 & 0.06 & 0.04 & 0.04 & 0.04\\
\hline
Normal(m=0, sd=1) & 0.46 & 0.38 & 0.42 & 0.64 & 0.53 & 0.59 & 0.46 & 0.39 & 0.44\\
\hline
Weibull(scale=1, shape=2) & 0.22 & 0.18 & 0.20 & 0.29 & 0.25 & 0.27 & 0.21 & 0.18 & 0.20\\
\hline
Student(df=3) & 0.69 & 0.63 & 0.65 & 0.81 & 0.76 & 0.76 & 0.59 & 0.55 & 0.56\\
\hline
Gumbel(loc=0, scale=1) & 0.54 & 0.48 & 0.51 & 0.69 & 0.63 & 0.66 & 0.50 & 0.46 & 0.48\\
\hline
Exp(rate=1) & 0.44 & 0.43 & 0.42 & 0.54 & 0.54 & 0.54 & 0.39 & 0.39 & 0.38\\
\hline
Cauchy(x0=0, gamma=1) & 1.73 & 15.51 & 1.94 & 1.41 & 2.21 & 1.47 & 1.00 & 1.40 & 1.02\\
\hline
Pareto(loc=1, shape=0.5) & 51.23 & 12909.42 & 72.10 & 7.91 & 55.16 & 13.44 & 4.18 & 17.87 & 6.23\\
\hline
Pareto(loc=1, shape=2) & 0.46 & 0.67 & 0.49 & 0.51 & 0.63 & 0.53 & 0.35 & 0.43 & 0.36\\
\hline
LogNormal(mlog=0, sdlog=1) & 0.76 & 0.87 & 0.77 & 0.75 & 0.88 & 0.81 & 0.52 & 0.61 & 0.54\\
\hline
LogNormal(mlog=0, sdlog=2) & 2.65 & 11.50 & 3.09 & 1.97 & 4.05 & 2.55 & 1.30 & 2.32 & 1.52\\
\hline
LogNormal(mlog=0, sdlog=3) & 14.29 & 192.92 & 24.39 & 4.74 & 20.40 & 7.43 & 2.35 & 9.98 & 3.83\\
\hline
Weibull(shape=0.3) & 4.17 & 11.45 & 5.16 & 1.67 & 5.81 & 2.89 & 0.80 & 3.00 & 1.36\\
\hline
Weibull(shape=0.5) & 1.23 & 1.72 & 1.34 & 1.14 & 1.72 & 1.31 & 0.70 & 1.08 & 0.83\\
\hline
Frechet(shape=1) & 2.03 & 10.32 & 2.30 & 1.50 & 3.10 & 1.79 & 0.99 & 1.74 & 1.13\\
\hline
Frechet(shape=3) & 0.28 & 0.30 & 0.28 & 0.31 & 0.31 & 0.30 & 0.22 & 0.22 & 0.22\\
\hline
\end{tabular}
\end{table}

\clearpage

\begin{table}[!h]

\caption{\label{tab:robust10}Properties of MAD estimations for n=10.}
\centering
\begin{tabular}[t]{l|r|r|r|r|r|r|r|r|r}
\hline
\multicolumn{1}{c|}{ } & \multicolumn{3}{c|}{SD} & \multicolumn{3}{c|}{IQR} & \multicolumn{3}{c}{MAD} \\
\cline{2-4} \cline{5-7} \cline{8-10}
Distribution & SM & HD & THD & SM & HD & THD & SM & HD & THD\\
\hline
Uniform(a=0, b=1) & 0.11 & 0.09 & 0.11 & 0.16 & 0.14 & 0.15 & 0.12 & 0.10 & 0.11\\
\hline
Triangular(a=0, b=2, c=1) & 0.16 & 0.14 & 0.15 & 0.21 & 0.18 & 0.19 & 0.16 & 0.13 & 0.14\\
\hline
Triangular(a=0, b=2, c=0.2) & 0.17 & 0.15 & 0.16 & 0.24 & 0.21 & 0.22 & 0.18 & 0.16 & 0.17\\
\hline
Beta(a=2, b=4) & 0.06 & 0.06 & 0.06 & 0.09 & 0.08 & 0.08 & 0.07 & 0.06 & 0.06\\
\hline
Beta(a=2, b=10) & 0.04 & 0.03 & 0.04 & 0.05 & 0.04 & 0.05 & 0.04 & 0.03 & 0.04\\
\hline
Normal(m=0, sd=1) & 0.37 & 0.32 & 0.35 & 0.51 & 0.43 & 0.47 & 0.38 & 0.32 & 0.34\\
\hline
Weibull(scale=1, shape=2) & 0.17 & 0.15 & 0.16 & 0.23 & 0.20 & 0.21 & 0.17 & 0.15 & 0.16\\
\hline
Student(df=3) & 0.52 & 0.47 & 0.49 & 0.66 & 0.60 & 0.62 & 0.48 & 0.43 & 0.45\\
\hline
Gumbel(loc=0, scale=1) & 0.42 & 0.38 & 0.41 & 0.58 & 0.53 & 0.55 & 0.43 & 0.39 & 0.40\\
\hline
Exp(rate=1) & 0.32 & 0.31 & 0.31 & 0.43 & 0.40 & 0.41 & 0.31 & 0.29 & 0.30\\
\hline
Cauchy(x0=0, gamma=1) & 0.99 & 1.46 & 1.01 & 1.12 & 1.24 & 1.08 & 0.79 & 0.88 & 0.76\\
\hline
Pareto(loc=1, shape=0.5) & 17.81 & 275359.25 & 36.33 & 5.99 & 21.77 & 7.51 & 3.67 & 10.20 & 4.53\\
\hline
Pareto(loc=1, shape=2) & 0.29 & 0.32 & 0.29 & 0.34 & 0.37 & 0.35 & 0.25 & 0.27 & 0.25\\
\hline
LogNormal(mlog=0, sdlog=1) & 0.49 & 0.51 & 0.49 & 0.57 & 0.62 & 0.58 & 0.41 & 0.45 & 0.43\\
\hline
LogNormal(mlog=0, sdlog=2) & 1.75 & 2.68 & 1.98 & 1.65 & 2.28 & 1.78 & 1.07 & 1.56 & 1.18\\
\hline
LogNormal(mlog=0, sdlog=3) & 5.23 & 18.53 & 6.12 & 2.77 & 7.80 & 3.69 & 1.60 & 4.72 & 2.21\\
\hline
Weibull(shape=0.3) & 1.97 & 3.57 & 2.44 & 1.09 & 2.84 & 1.59 & 0.59 & 1.73 & 0.90\\
\hline
Weibull(shape=0.5) & 0.83 & 0.93 & 0.84 & 0.89 & 1.08 & 0.94 & 0.60 & 0.74 & 0.63\\
\hline
Frechet(shape=1) & 1.07 & 2.21 & 1.19 & 1.04 & 1.47 & 1.13 & 0.74 & 1.02 & 0.76\\
\hline
Frechet(shape=3) & 0.19 & 0.19 & 0.19 & 0.25 & 0.24 & 0.24 & 0.18 & 0.17 & 0.18\\
\hline
\end{tabular}
\end{table}

\begin{table}[!h]

\caption{\label{tab:robust50}Properties of MAD estimations for n=50.}
\centering
\begin{tabular}[t]{l|r|r|r|r|r|r|r|r|r}
\hline
\multicolumn{1}{c|}{ } & \multicolumn{3}{c|}{SD} & \multicolumn{3}{c|}{IQR} & \multicolumn{3}{c}{MAD} \\
\cline{2-4} \cline{5-7} \cline{8-10}
Distribution & SM & HD & THD & SM & HD & THD & SM & HD & THD\\
\hline
Uniform(a=0, b=1) & 0.05 & 0.05 & 0.05 & 0.07 & 0.07 & 0.07 & 0.05 & 0.05 & 0.05\\
\hline
Triangular(a=0, b=2, c=1) & 0.07 & 0.07 & 0.07 & 0.11 & 0.09 & 0.10 & 0.08 & 0.07 & 0.07\\
\hline
Triangular(a=0, b=2, c=0.2) & 0.08 & 0.08 & 0.08 & 0.11 & 0.10 & 0.11 & 0.08 & 0.08 & 0.08\\
\hline
Beta(a=2, b=4) & 0.03 & 0.03 & 0.03 & 0.04 & 0.04 & 0.04 & 0.03 & 0.03 & 0.03\\
\hline
Beta(a=2, b=10) & 0.02 & 0.02 & 0.02 & 0.02 & 0.02 & 0.02 & 0.02 & 0.02 & 0.02\\
\hline
Normal(m=0, sd=1) & 0.17 & 0.16 & 0.16 & 0.22 & 0.21 & 0.22 & 0.16 & 0.16 & 0.16\\
\hline
Weibull(scale=1, shape=2) & 0.07 & 0.07 & 0.07 & 0.10 & 0.09 & 0.09 & 0.07 & 0.07 & 0.07\\
\hline
Student(df=3) & 0.21 & 0.20 & 0.20 & 0.28 & 0.26 & 0.26 & 0.21 & 0.19 & 0.20\\
\hline
Gumbel(loc=0, scale=1) & 0.20 & 0.19 & 0.19 & 0.26 & 0.24 & 0.25 & 0.19 & 0.18 & 0.19\\
\hline
Exp(rate=1) & 0.15 & 0.14 & 0.14 & 0.19 & 0.18 & 0.19 & 0.14 & 0.14 & 0.14\\
\hline
Cauchy(x0=0, gamma=1) & 0.36 & 0.34 & 0.35 & 0.46 & 0.44 & 0.44 & 0.33 & 0.33 & 0.32\\
\hline
Pareto(loc=1, shape=0.5) & 1.83 & 2.02 & 1.87 & 2.05 & 2.25 & 2.04 & 1.42 & 1.55 & 1.48\\
\hline
Pareto(loc=1, shape=2) & 0.12 & 0.12 & 0.12 & 0.16 & 0.15 & 0.15 & 0.11 & 0.11 & 0.11\\
\hline
LogNormal(mlog=0, sdlog=1) & 0.20 & 0.19 & 0.20 & 0.28 & 0.27 & 0.27 & 0.20 & 0.20 & 0.20\\
\hline
LogNormal(mlog=0, sdlog=2) & 0.53 & 0.54 & 0.52 & 0.66 & 0.65 & 0.64 & 0.47 & 0.48 & 0.48\\
\hline
LogNormal(mlog=0, sdlog=3) & 0.97 & 1.05 & 0.96 & 1.01 & 1.20 & 1.05 & 0.71 & 0.82 & 0.74\\
\hline
Weibull(shape=0.3) & 0.38 & 0.42 & 0.38 & 0.40 & 0.46 & 0.42 & 0.28 & 0.33 & 0.29\\
\hline
Weibull(shape=0.5) & 0.29 & 0.28 & 0.28 & 0.35 & 0.35 & 0.34 & 0.26 & 0.25 & 0.25\\
\hline
Frechet(shape=1) & 0.40 & 0.41 & 0.40 & 0.50 & 0.48 & 0.47 & 0.36 & 0.34 & 0.34\\
\hline
Frechet(shape=3) & 0.08 & 0.08 & 0.08 & 0.11 & 0.11 & 0.11 & 0.08 & 0.08 & 0.08\\
\hline
\end{tabular}
\end{table}

\clearpage

\clearpage

\hypertarget{sec-cases}{%
\section{Special cases of bias-correction factors}\label{sec-cases}}

In this section, we consider two following cases:

\begin{itemize}
\tightlist
\item
  \(n=2\): it is the only case when we can easily calculate the exact value of the bias correction factor.
\item
  \(n>100\): for this case, we draw a generic equation following the approach from \autocite{hayes2014}.
\end{itemize}

\hypertarget{bias-correction-factors-for-n-2}{%
\subsection{Bias-correction factors for n = 2}\label{bias-correction-factors-for-n-2}}

Let \(X = \{ X_1, X_2 \}\) be a sample of two i.i.d. random variables
from the standard normal distribution \(\mathcal{N}(0, 1^2)\).
Regardless of the chosen median estimator, the median is unequivocally determined:

\[
\operatorname{median}(X) = \dfrac{X_1 + X_2}{2}.
\]

Now we calculate the median absolute deviation \(\operatorname{MAD}_0\):

\[
\begin{split}
\operatorname{MAD}_0(X)
  & = \operatorname{median}(|X - \operatorname{median}(X)|) = \\
  & = \operatorname{median}(\{ \, |X_1 - (X_1 + X_2)/2|\,,\, |X_2 - (X_1 + X_2)/2|\, \}) = \\
  & = \operatorname{median}(\{ \, |(X_1 - X_2)/2|\,,\, |(X_2 - X_1)/2| \,\}) = \\
  & = |X_1 - X_2|/2.
\end{split}
\]

Since \(X_1, X_2 \sim \mathcal{N}(0, 1^2)\) which is symmetric,
\(|X_1 - X_2|/2\) is distributed the same way as \(|X_1 + X_2|/2\).
Let us denote the sum of two standard normal distributions by \(Z = X_1 + X_2\).
It gives us another normal distribution with modified variance:

\[
Z \sim \mathcal{N}(0, \sqrt{2}^2).
\]

Since we take the absolute value of \(Z\), we get the half-normal distribution.
The expected value of a half-normal distribution which is formed from the normal distribution \(\mathcal{N}(0, \sigma^2)\)
is \(\sigma \sqrt{2/\pi}\).
Thus,

\[
\mathbb{E}[|Z|] = \sqrt{2} \sqrt{2/\pi} = 2/\sqrt{\pi}.
\]

Finally, we have:

\[
\mathbb{E}[\operatorname{MAD}_0(X)]
  = \mathbb{E}\Bigg[ \frac{|X_1 - X_2|}{2} \Bigg]
  = \mathbb{E}\Bigg[ \frac{|X_1 + X_2|}{2} \Bigg]
  = \mathbb{E}\Bigg[ \frac{|Z|}{2} \Bigg]
  = \frac{2/\sqrt{\pi}}{2} = \frac{1}{\sqrt{\pi}}.
\]

The bias-correction factor \(C_2\) is the reciprocal value of the expected value of \(\operatorname{MAD}_0(X)\):

\[
C_2 = \frac{1}{\mathbb{E}[\operatorname{MAD}_0(X)]} = \sqrt{\pi} \approx 1.77245385090552.
\]

\clearpage

\hypertarget{bias-correction-factors-for-n-100}{%
\subsection{Bias-correction factors for n \textgreater{} 100}\label{bias-correction-factors-for-n-100}}

Following the approach from \autocite{hayes2014}, we are going to draw a generic equation for \(C_n\) in the following form:

\[
C_n = \dfrac{1}{\Phi^{-1}(0.75) \cdot (1+A_n)}, \quad A_n = \dfrac{\alpha}{n} + \dfrac{\beta}{n^2}.
\]

The coefficients \(\alpha\) and \(\beta\) can be obtained using least squares
on the values from Tables~\ref{tab:park}
(let us denote \(\operatorname{MAD}\) based on this table by \(\operatorname{MAD}_{\operatorname{PARK}}\)),
\ref{tab:sm}, \ref{tab:hd}, and \ref{tab:thd-sqrt} for \(100 < n \leq 500\).
The results are presented in Table~\ref{tab:ab}.

\begin{table}[!h]

\caption{\label{tab:ab}$A_n$ parameters for $n > 100$.}
\centering
\begin{tabular}[t]{l|r|r}
\hline
  & $\alpha$ & $\beta$\\
\hline
$\operatorname{MAD}_{\operatorname{PARK}}$ & -0.7591 & -1.3239\\
\hline
$\operatorname{MAD}_{\operatorname{SM}}$ & -0.7668 & -2.1897\\
\hline
$\operatorname{MAD}_{\operatorname{HD}}$ & -0.4912 & -7.6350\\
\hline
$\operatorname{MAD}_{\operatorname{THD-SQRT}}$ & -0.6954 & -4.9261\\
\hline
\end{tabular}
\end{table}

The value of \(\alpha\) for \(\operatorname{MAD}_{\operatorname{PARK}}\) and \(\operatorname{MAD}_{\operatorname{SM}}\)
are quite close to the suggested \(\alpha=-0.76213\) from \autocite{park2020}.
The corresponding value of \(\beta\) is not so close to \(\beta=-0.86413\) from \autocite{park2020},
but this difference does not produce a noticeable impact on the final result.

\begin{figure}[ht!]

{\centering \includegraphics{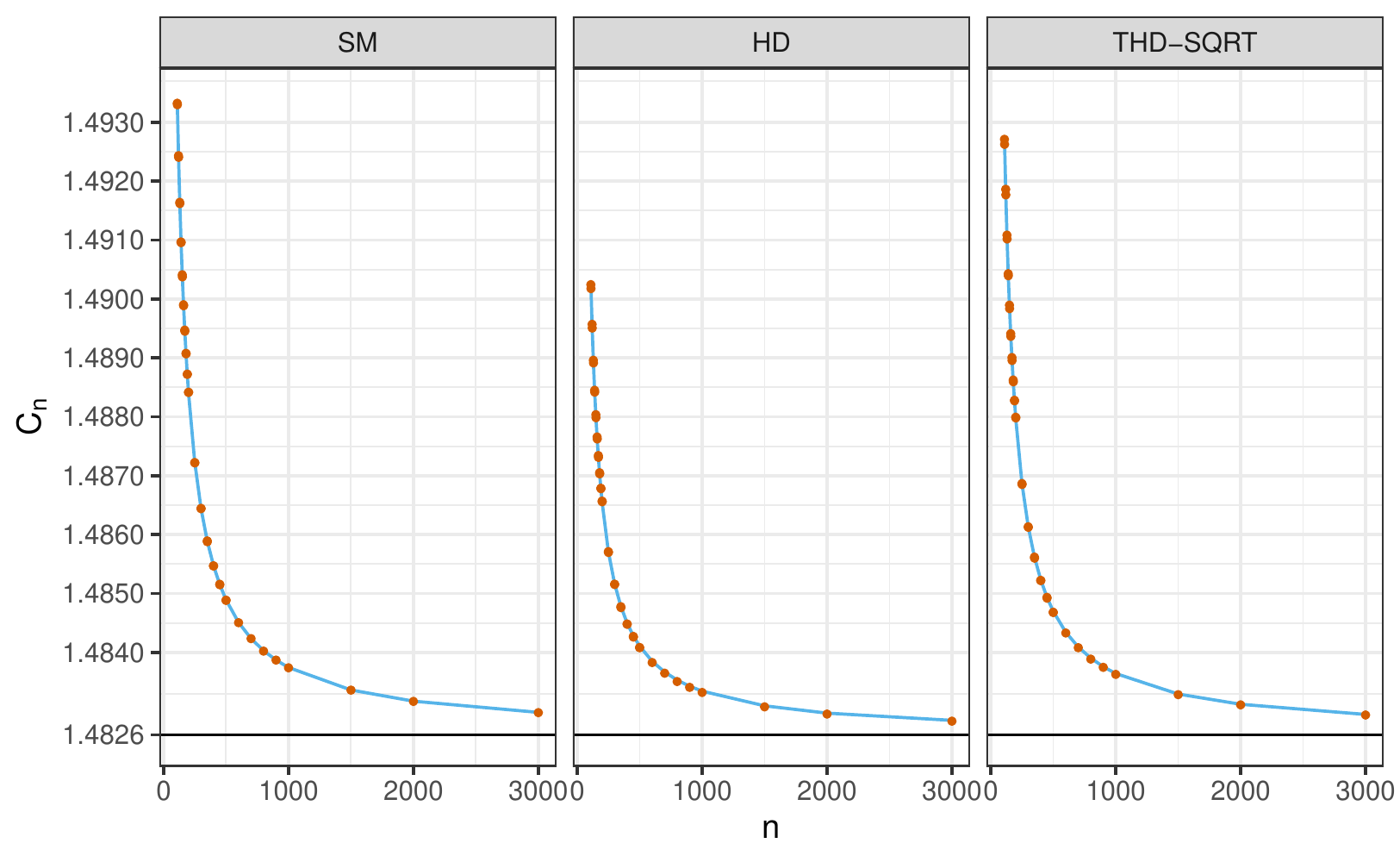} 

}

\caption{Actual and predicted bias-correction factors}\label{fig:n100}
\end{figure}

The evaluated values of \(\alpha\) and \(\beta\) for all \(\operatorname{MAD}\) estimators look quite accurate.
In Figure~\ref{fig:n100}, we can see the actual (points) and predicted (line) values of \(C_n\) for \(100 < n \leq 3000\).
Within values \(500 < n \leq 3000\) from Tables~\ref{tab:sm}, \ref{tab:hd}, and \ref{tab:thd-sqrt}
(that were not used to get the values of \(\alpha\) and \(\beta\)),
the maximum observed absolute difference between the actual and predicted values
is \(\approx 0.000061\).

\clearpage

\hypertarget{sec-summary}{%
\section{Summary}\label{sec-summary}}

The median absolute deviation is a robust measure of statistical dispersion
that can be used as a consistent estimator for the standard deviation
under the normal distribution.
To make it unbiased, we have to use a bias-correction factor \(C_n\):

\[
\operatorname{MAD}_n(X) = C_n \cdot \operatorname{median}(|X-\operatorname{median}(X)|).
\]

This approach heavily depends on the chosen median estimator.
In this paper, we have discussed three estimators:
the classic sample median (\(\operatorname{median}_{\operatorname{SM}}\)),
the Harrell-Davis quantile estimator (\(\operatorname{median}_{\operatorname{HD}}\)),
and the trimmed Harrell-Davis quantile estimator
based on the highest density interval of the width \(1/\sqrt{n}\) (\(\operatorname{median}_{\operatorname{THD-SQRT}}\))
which give us estimators
\(\operatorname{MAD}_{\operatorname{SM}}\) and \(\operatorname{MAD}_{\operatorname{HD}}\),
and \(\operatorname{MAD}_{\operatorname{THD-SQRT}}\) respectively.

In Simulation~\protect\hyperlink{sim1}{1}, we estimated values of \(C_n\) using the Monte-Carlo simulation for each estimator.
These values are listed in Tables~\ref{tab:sm}, \ref{tab:hd}, and \ref{tab:thd-sqrt}.
These tables cover all values of \(n\) from \(2\) to \(100\) and some greater values up to \(3000\).
A generic approach for large sample sizes (\(n > 100\)) can be presented in the following form:

\[
C_n = \dfrac{1}{\Phi^{-1}(0.75) \cdot (1+\alpha/n + \beta/n^2)},
\]

where the values of \(\alpha\) and \(\beta\) are listed in Table \ref{tab:ab}.
For \(n=2\), we know the exact value of the bias-correction factor:
\(C_2 = \sqrt{\pi} \approx 1.77245385090552\).

In Simulation~\protect\hyperlink{sim2}{2}, we evaluated the relative statistical efficiency of
\(\operatorname{MAD}_{\operatorname{HD}}\) and \(\operatorname{MAD}_{\operatorname{THD-SQRT}}\)
against \(\operatorname{MAD}_{\operatorname{SM}}\).
It turned out that the efficiency of
\(\operatorname{MAD}_{\operatorname{HD}}\) and \(\operatorname{MAD}_{\operatorname{THD-SQRT}}\)
are noticeably higher than the efficiency of \(\operatorname{MAD}_{\operatorname{SM}}\).

In Simulation~\protect\hyperlink{sim3}{3}, we investigated the sensitivity to outliers of all \(\operatorname{MAD}\) estimators.
It turned out that \(\operatorname{MAD}_{\operatorname{HD}}\) could be corrupted by extreme outliers
in the case of heavy-tailed distributions.
Meanwhile, \(\operatorname{MAD}_{\operatorname{THD-SQRT}}\) is much more resistant to outliers
(while it is still not as robust as \(\operatorname{MAD}_{\operatorname{SM}}\)).

Thus, in the case of light-tailed distributions,
we recommend \(\operatorname{MAD}_{\operatorname{HD}}\) as an alternative
to the classic \(\operatorname{MAD}_{\operatorname{SM}}\)
because it has higher statistical efficiency.
In the case of heavy-tailed distributions,
we recommend \(\operatorname{MAD}_{\operatorname{THD-SQRT}}\)
because it allows achieving a good trade-off between statistical efficiency and robustness.
The practical impact of both approaches is most noticeable for samples with a small number of elements.

The trade-off between the statistical efficiency and the robustness
can be customized by choosing another width of the Beta distribution's highest density interval
in \(\operatorname{MAD}_{\operatorname{THD}}\).
The exact value of this width should be carefully chosen based on
the knowledge of the considered distribution,
the expected number and the magnitude of possible outliers,
and the robustness requirements.
The values of the bias-correction factors \(C_n\) should be properly updated
using another Monte-Carlo simulation study similar to Simulation \protect\hyperlink{sim1}{1}.

\hypertarget{disclosure-statement}{%
\section*{Disclosure statement}\label{disclosure-statement}}
\addcontentsline{toc}{section}{Disclosure statement}

The author reports there are no competing interests to declare.

\hypertarget{data-and-source-code-availability}{%
\section*{Data and source code availability}\label{data-and-source-code-availability}}
\addcontentsline{toc}{section}{Data and source code availability}

The source code of this paper, the source code of all simulations,
and the simulation results are available on GitHub:
\url{https://github.com/AndreyAkinshin/paper-mad-factors}.

\hypertarget{acknowledgments}{%
\section*{Acknowledgments}\label{acknowledgments}}
\addcontentsline{toc}{section}{Acknowledgments}

The author thanks Ivan Pashchenko for valuable discussions.

\clearpage

\hypertarget{sec-ref}{%
\section*{A~~Reference implementation}\label{sec-ref}}
\addcontentsline{toc}{section}{A~~Reference implementation}

Here is an R implementation of the suggested \(\operatorname{MAD}\) estimators:

\begin{Shaded}
\begin{Highlighting}[]
\NormalTok{quantile.hd }\OtherTok{\textless{}{-}} \ControlFlowTok{function}\NormalTok{(x, probs) }\FunctionTok{sapply}\NormalTok{(probs, }\ControlFlowTok{function}\NormalTok{(p) \{}
\NormalTok{  n }\OtherTok{\textless{}{-}} \FunctionTok{length}\NormalTok{(x)}
  \ControlFlowTok{if}\NormalTok{ (n }\SpecialCharTok{==} \DecValTok{0}\NormalTok{) }\FunctionTok{return}\NormalTok{(}\ConstantTok{NA}\NormalTok{)}
  \ControlFlowTok{if}\NormalTok{ (n }\SpecialCharTok{==} \DecValTok{1}\NormalTok{) }\FunctionTok{return}\NormalTok{(x)}
\NormalTok{  x }\OtherTok{\textless{}{-}} \FunctionTok{sort}\NormalTok{(x)}
\NormalTok{  a }\OtherTok{\textless{}{-}}\NormalTok{ (n }\SpecialCharTok{+} \DecValTok{1}\NormalTok{) }\SpecialCharTok{*}\NormalTok{ p; b }\OtherTok{\textless{}{-}}\NormalTok{ (n }\SpecialCharTok{+} \DecValTok{1}\NormalTok{) }\SpecialCharTok{*}\NormalTok{ (}\DecValTok{1} \SpecialCharTok{{-}}\NormalTok{ p)}
\NormalTok{  cdfs }\OtherTok{\textless{}{-}} \FunctionTok{pbeta}\NormalTok{(}\DecValTok{0}\SpecialCharTok{:}\NormalTok{n}\SpecialCharTok{/}\NormalTok{n, a, b)}
\NormalTok{  W }\OtherTok{\textless{}{-}} \FunctionTok{tail}\NormalTok{(cdfs, }\SpecialCharTok{{-}}\DecValTok{1}\NormalTok{) }\SpecialCharTok{{-}} \FunctionTok{head}\NormalTok{(cdfs, }\SpecialCharTok{{-}}\DecValTok{1}\NormalTok{)}
  \FunctionTok{sum}\NormalTok{(x }\SpecialCharTok{*}\NormalTok{ W)}
\NormalTok{\})}
\NormalTok{quantile.thd }\OtherTok{\textless{}{-}} \ControlFlowTok{function}\NormalTok{(x, probs, }\AttributeTok{width =} \DecValTok{1}\SpecialCharTok{/}\FunctionTok{sqrt}\NormalTok{(}\FunctionTok{length}\NormalTok{(x))) }\FunctionTok{sapply}\NormalTok{(probs, }\ControlFlowTok{function}\NormalTok{(p) \{}
\NormalTok{  getBetaHdi }\OtherTok{\textless{}{-}} \ControlFlowTok{function}\NormalTok{(a, b, width) \{}
\NormalTok{    eps }\OtherTok{\textless{}{-}} \FloatTok{1e{-}9}
    \ControlFlowTok{if}\NormalTok{ (a }\SpecialCharTok{\textless{}} \DecValTok{1} \SpecialCharTok{+}\NormalTok{ eps }\SpecialCharTok{\&}\NormalTok{ b }\SpecialCharTok{\textless{}} \DecValTok{1} \SpecialCharTok{+}\NormalTok{ eps) }\CommentTok{\# Degenerate case}
      \FunctionTok{return}\NormalTok{(}\FunctionTok{c}\NormalTok{(}\ConstantTok{NA}\NormalTok{, }\ConstantTok{NA}\NormalTok{))}
    \ControlFlowTok{if}\NormalTok{ (a }\SpecialCharTok{\textless{}} \DecValTok{1} \SpecialCharTok{+}\NormalTok{ eps }\SpecialCharTok{\&}\NormalTok{ b }\SpecialCharTok{\textgreater{}} \DecValTok{1}\NormalTok{) }\CommentTok{\# Left border case}
      \FunctionTok{return}\NormalTok{(}\FunctionTok{c}\NormalTok{(}\DecValTok{0}\NormalTok{, width))}
    \ControlFlowTok{if}\NormalTok{ (a }\SpecialCharTok{\textgreater{}} \DecValTok{1} \SpecialCharTok{\&}\NormalTok{ b }\SpecialCharTok{\textless{}} \DecValTok{1} \SpecialCharTok{+}\NormalTok{ eps) }\CommentTok{\# Right border case}
      \FunctionTok{return}\NormalTok{(}\FunctionTok{c}\NormalTok{(}\DecValTok{1} \SpecialCharTok{{-}}\NormalTok{ width, }\DecValTok{1}\NormalTok{))}
    \ControlFlowTok{if}\NormalTok{ (width }\SpecialCharTok{\textgreater{}} \DecValTok{1} \SpecialCharTok{{-}}\NormalTok{ eps)}
      \FunctionTok{return}\NormalTok{(}\FunctionTok{c}\NormalTok{(}\DecValTok{0}\NormalTok{, }\DecValTok{1}\NormalTok{))}
    \CommentTok{\# Middle case}
\NormalTok{    mode }\OtherTok{\textless{}{-}}\NormalTok{ (a }\SpecialCharTok{{-}} \DecValTok{1}\NormalTok{) }\SpecialCharTok{/}\NormalTok{ (a }\SpecialCharTok{+}\NormalTok{ b }\SpecialCharTok{{-}} \DecValTok{2}\NormalTok{)}
\NormalTok{    pdf }\OtherTok{\textless{}{-}} \ControlFlowTok{function}\NormalTok{(x) }\FunctionTok{dbeta}\NormalTok{(x, a, b)}
\NormalTok{    l }\OtherTok{\textless{}{-}} \FunctionTok{uniroot}\NormalTok{(}
      \AttributeTok{f =} \ControlFlowTok{function}\NormalTok{(x) }\FunctionTok{pdf}\NormalTok{(x) }\SpecialCharTok{{-}} \FunctionTok{pdf}\NormalTok{(x }\SpecialCharTok{+}\NormalTok{ width),}
      \AttributeTok{lower =} \FunctionTok{max}\NormalTok{(}\DecValTok{0}\NormalTok{, mode }\SpecialCharTok{{-}}\NormalTok{ width),}
      \AttributeTok{upper =} \FunctionTok{min}\NormalTok{(mode, }\DecValTok{1} \SpecialCharTok{{-}}\NormalTok{ width),}
      \AttributeTok{tol =} \FloatTok{1e{-}9}
\NormalTok{    )}\SpecialCharTok{$}\NormalTok{root}
\NormalTok{    r }\OtherTok{\textless{}{-}}\NormalTok{ l }\SpecialCharTok{+}\NormalTok{ width}
    \FunctionTok{return}\NormalTok{(}\FunctionTok{c}\NormalTok{(l, r))}
\NormalTok{  \}}
\NormalTok{  n }\OtherTok{\textless{}{-}} \FunctionTok{length}\NormalTok{(x)}
  \ControlFlowTok{if}\NormalTok{ (n }\SpecialCharTok{==} \DecValTok{0}\NormalTok{) }\FunctionTok{return}\NormalTok{(}\ConstantTok{NA}\NormalTok{)}
  \ControlFlowTok{if}\NormalTok{ (n }\SpecialCharTok{==} \DecValTok{1}\NormalTok{) }\FunctionTok{return}\NormalTok{(x)}
\NormalTok{  x }\OtherTok{\textless{}{-}} \FunctionTok{sort}\NormalTok{(x)}
\NormalTok{  a }\OtherTok{\textless{}{-}}\NormalTok{ (n }\SpecialCharTok{+} \DecValTok{1}\NormalTok{) }\SpecialCharTok{*}\NormalTok{ p; b }\OtherTok{\textless{}{-}}\NormalTok{ (n }\SpecialCharTok{+} \DecValTok{1}\NormalTok{) }\SpecialCharTok{*}\NormalTok{ (}\DecValTok{1} \SpecialCharTok{{-}}\NormalTok{ p)}
\NormalTok{  hdi }\OtherTok{\textless{}{-}} \FunctionTok{getBetaHdi}\NormalTok{(a, b, width)}
\NormalTok{  hdiCdf }\OtherTok{\textless{}{-}} \FunctionTok{pbeta}\NormalTok{(hdi, a, b)}
\NormalTok{  cdf }\OtherTok{\textless{}{-}} \ControlFlowTok{function}\NormalTok{(xs) \{}
\NormalTok{    xs[xs }\SpecialCharTok{\textless{}=}\NormalTok{ hdi[}\DecValTok{1}\NormalTok{]] }\OtherTok{\textless{}{-}}\NormalTok{ hdi[}\DecValTok{1}\NormalTok{]}
\NormalTok{    xs[xs }\SpecialCharTok{\textgreater{}=}\NormalTok{ hdi[}\DecValTok{2}\NormalTok{]] }\OtherTok{\textless{}{-}}\NormalTok{ hdi[}\DecValTok{2}\NormalTok{]}
\NormalTok{    (}\FunctionTok{pbeta}\NormalTok{(xs, a, b) }\SpecialCharTok{{-}}\NormalTok{ hdiCdf[}\DecValTok{1}\NormalTok{]) }\SpecialCharTok{/}\NormalTok{ (hdiCdf[}\DecValTok{2}\NormalTok{] }\SpecialCharTok{{-}}\NormalTok{ hdiCdf[}\DecValTok{1}\NormalTok{])}
\NormalTok{  \}}
\NormalTok{  iL }\OtherTok{\textless{}{-}} \FunctionTok{floor}\NormalTok{(hdi[}\DecValTok{1}\NormalTok{] }\SpecialCharTok{*}\NormalTok{ n); iR }\OtherTok{\textless{}{-}} \FunctionTok{ceiling}\NormalTok{(hdi[}\DecValTok{2}\NormalTok{] }\SpecialCharTok{*}\NormalTok{ n)}
\NormalTok{  cdfs }\OtherTok{\textless{}{-}} \FunctionTok{cdf}\NormalTok{(iL}\SpecialCharTok{:}\NormalTok{iR}\SpecialCharTok{/}\NormalTok{n)}
\NormalTok{  W }\OtherTok{\textless{}{-}} \FunctionTok{tail}\NormalTok{(cdfs, }\SpecialCharTok{{-}}\DecValTok{1}\NormalTok{) }\SpecialCharTok{{-}} \FunctionTok{head}\NormalTok{(cdfs, }\SpecialCharTok{{-}}\DecValTok{1}\NormalTok{)}
  \FunctionTok{sum}\NormalTok{(x[(iL }\SpecialCharTok{+} \DecValTok{1}\NormalTok{)}\SpecialCharTok{:}\NormalTok{iR] }\SpecialCharTok{*}\NormalTok{ W)}
\NormalTok{\})}
\NormalTok{med.sm }\OtherTok{\textless{}{-}} \ControlFlowTok{function}\NormalTok{(x) }\FunctionTok{median}\NormalTok{(x)}
\NormalTok{med.hd }\OtherTok{\textless{}{-}} \ControlFlowTok{function}\NormalTok{(x) }\FunctionTok{quantile.hd}\NormalTok{(x, }\FloatTok{0.5}\NormalTok{)}
\NormalTok{med.thd.sqrt }\OtherTok{\textless{}{-}} \ControlFlowTok{function}\NormalTok{(x) }\FunctionTok{quantile.thd}\NormalTok{(x, }\FloatTok{0.5}\NormalTok{)}

\NormalTok{factors.sm }\OtherTok{\textless{}{-}} \FunctionTok{c}\NormalTok{(}
      \ConstantTok{NA}\NormalTok{, }\FloatTok{1.7725}\NormalTok{, }\FloatTok{2.2049}\NormalTok{, }\FloatTok{2.0172}\NormalTok{, }\FloatTok{1.8040}\NormalTok{, }\FloatTok{1.7637}\NormalTok{, }\FloatTok{1.6871}\NormalTok{, }\FloatTok{1.6715}\NormalTok{, }\FloatTok{1.6326}\NormalTok{, }\FloatTok{1.6245}\NormalTok{,}
  \FloatTok{1.6011}\NormalTok{, }\FloatTok{1.5961}\NormalTok{, }\FloatTok{1.5806}\NormalTok{, }\FloatTok{1.5772}\NormalTok{, }\FloatTok{1.5661}\NormalTok{, }\FloatTok{1.5637}\NormalTok{, }\FloatTok{1.5554}\NormalTok{, }\FloatTok{1.5536}\NormalTok{, }\FloatTok{1.5471}\NormalTok{, }\FloatTok{1.5457}\NormalTok{,}
  \FloatTok{1.5405}\NormalTok{, }\FloatTok{1.5393}\NormalTok{, }\FloatTok{1.5352}\NormalTok{, }\FloatTok{1.5342}\NormalTok{, }\FloatTok{1.5307}\NormalTok{, }\FloatTok{1.5299}\NormalTok{, }\FloatTok{1.5269}\NormalTok{, }\FloatTok{1.5263}\NormalTok{, }\FloatTok{1.5238}\NormalTok{, }\FloatTok{1.5233}\NormalTok{,}
  \FloatTok{1.5212}\NormalTok{, }\FloatTok{1.5207}\NormalTok{, }\FloatTok{1.5189}\NormalTok{, }\FloatTok{1.5184}\NormalTok{, }\FloatTok{1.5168}\NormalTok{, }\FloatTok{1.5164}\NormalTok{, }\FloatTok{1.5149}\NormalTok{, }\FloatTok{1.5146}\NormalTok{, }\FloatTok{1.5132}\NormalTok{, }\FloatTok{1.5129}\NormalTok{,}
  \FloatTok{1.5117}\NormalTok{, }\FloatTok{1.5115}\NormalTok{, }\FloatTok{1.5103}\NormalTok{, }\FloatTok{1.5101}\NormalTok{, }\FloatTok{1.5091}\NormalTok{, }\FloatTok{1.5089}\NormalTok{, }\FloatTok{1.5080}\NormalTok{, }\FloatTok{1.5078}\NormalTok{, }\FloatTok{1.5069}\NormalTok{, }\FloatTok{1.5067}\NormalTok{,}
  \FloatTok{1.5060}\NormalTok{, }\FloatTok{1.5058}\NormalTok{, }\FloatTok{1.5051}\NormalTok{, }\FloatTok{1.5049}\NormalTok{, }\FloatTok{1.5042}\NormalTok{, }\FloatTok{1.5041}\NormalTok{, }\FloatTok{1.5035}\NormalTok{, }\FloatTok{1.5033}\NormalTok{, }\FloatTok{1.5027}\NormalTok{, }\FloatTok{1.5026}\NormalTok{,}
  \FloatTok{1.5021}\NormalTok{, }\FloatTok{1.5019}\NormalTok{, }\FloatTok{1.5014}\NormalTok{, }\FloatTok{1.5013}\NormalTok{, }\FloatTok{1.5008}\NormalTok{, }\FloatTok{1.5007}\NormalTok{, }\FloatTok{1.5003}\NormalTok{, }\FloatTok{1.5002}\NormalTok{, }\FloatTok{1.4998}\NormalTok{, }\FloatTok{1.4997}\NormalTok{,}
  \FloatTok{1.4993}\NormalTok{, }\FloatTok{1.4992}\NormalTok{, }\FloatTok{1.4988}\NormalTok{, }\FloatTok{1.4987}\NormalTok{, }\FloatTok{1.4984}\NormalTok{, }\FloatTok{1.4983}\NormalTok{, }\FloatTok{1.4979}\NormalTok{, }\FloatTok{1.4978}\NormalTok{, }\FloatTok{1.4975}\NormalTok{, }\FloatTok{1.4975}\NormalTok{,}
  \FloatTok{1.4972}\NormalTok{, }\FloatTok{1.4971}\NormalTok{, }\FloatTok{1.4968}\NormalTok{, }\FloatTok{1.4967}\NormalTok{, }\FloatTok{1.4965}\NormalTok{, }\FloatTok{1.4964}\NormalTok{, }\FloatTok{1.4961}\NormalTok{, }\FloatTok{1.4961}\NormalTok{, }\FloatTok{1.4958}\NormalTok{, }\FloatTok{1.4958}\NormalTok{,}
  \FloatTok{1.4955}\NormalTok{, }\FloatTok{1.4955}\NormalTok{, }\FloatTok{1.4952}\NormalTok{, }\FloatTok{1.4952}\NormalTok{, }\FloatTok{1.4950}\NormalTok{, }\FloatTok{1.4949}\NormalTok{, }\FloatTok{1.4947}\NormalTok{, }\FloatTok{1.4947}\NormalTok{, }\FloatTok{1.4945}\NormalTok{, }\FloatTok{1.4944}\NormalTok{)}
\NormalTok{factors.hd }\OtherTok{\textless{}{-}} \FunctionTok{c}\NormalTok{(}
      \ConstantTok{NA}\NormalTok{, }\FloatTok{1.7725}\NormalTok{, }\FloatTok{1.5682}\NormalTok{, }\FloatTok{1.5959}\NormalTok{, }\FloatTok{1.5661}\NormalTok{, }\FloatTok{1.5666}\NormalTok{, }\FloatTok{1.5646}\NormalTok{, }\FloatTok{1.5591}\NormalTok{, }\FloatTok{1.5567}\NormalTok{, }\FloatTok{1.5529}\NormalTok{,}
  \FloatTok{1.5496}\NormalTok{, }\FloatTok{1.5465}\NormalTok{, }\FloatTok{1.5434}\NormalTok{, }\FloatTok{1.5406}\NormalTok{, }\FloatTok{1.5380}\NormalTok{, }\FloatTok{1.5355}\NormalTok{, }\FloatTok{1.5332}\NormalTok{, }\FloatTok{1.5310}\NormalTok{, }\FloatTok{1.5289}\NormalTok{, }\FloatTok{1.5270}\NormalTok{,}
  \FloatTok{1.5252}\NormalTok{, }\FloatTok{1.5235}\NormalTok{, }\FloatTok{1.5220}\NormalTok{, }\FloatTok{1.5204}\NormalTok{, }\FloatTok{1.5191}\NormalTok{, }\FloatTok{1.5177}\NormalTok{, }\FloatTok{1.5164}\NormalTok{, }\FloatTok{1.5154}\NormalTok{, }\FloatTok{1.5143}\NormalTok{, }\FloatTok{1.5133}\NormalTok{,}
  \FloatTok{1.5123}\NormalTok{, }\FloatTok{1.5114}\NormalTok{, }\FloatTok{1.5106}\NormalTok{, }\FloatTok{1.5098}\NormalTok{, }\FloatTok{1.5090}\NormalTok{, }\FloatTok{1.5083}\NormalTok{, }\FloatTok{1.5076}\NormalTok{, }\FloatTok{1.5069}\NormalTok{, }\FloatTok{1.5062}\NormalTok{, }\FloatTok{1.5056}\NormalTok{,}
  \FloatTok{1.5050}\NormalTok{, }\FloatTok{1.5045}\NormalTok{, }\FloatTok{1.5039}\NormalTok{, }\FloatTok{1.5034}\NormalTok{, }\FloatTok{1.5029}\NormalTok{, }\FloatTok{1.5025}\NormalTok{, }\FloatTok{1.5020}\NormalTok{, }\FloatTok{1.5016}\NormalTok{, }\FloatTok{1.5011}\NormalTok{, }\FloatTok{1.5008}\NormalTok{,}
  \FloatTok{1.5004}\NormalTok{, }\FloatTok{1.5000}\NormalTok{, }\FloatTok{1.4997}\NormalTok{, }\FloatTok{1.4993}\NormalTok{, }\FloatTok{1.4990}\NormalTok{, }\FloatTok{1.4986}\NormalTok{, }\FloatTok{1.4983}\NormalTok{, }\FloatTok{1.4980}\NormalTok{, }\FloatTok{1.4977}\NormalTok{, }\FloatTok{1.4975}\NormalTok{,}
  \FloatTok{1.4972}\NormalTok{, }\FloatTok{1.4969}\NormalTok{, }\FloatTok{1.4967}\NormalTok{, }\FloatTok{1.4964}\NormalTok{, }\FloatTok{1.4962}\NormalTok{, }\FloatTok{1.4960}\NormalTok{, }\FloatTok{1.4957}\NormalTok{, }\FloatTok{1.4955}\NormalTok{, }\FloatTok{1.4953}\NormalTok{, }\FloatTok{1.4951}\NormalTok{,}
  \FloatTok{1.4950}\NormalTok{, }\FloatTok{1.4947}\NormalTok{, }\FloatTok{1.4946}\NormalTok{, }\FloatTok{1.4944}\NormalTok{, }\FloatTok{1.4942}\NormalTok{, }\FloatTok{1.4940}\NormalTok{, }\FloatTok{1.4939}\NormalTok{, }\FloatTok{1.4937}\NormalTok{, }\FloatTok{1.4936}\NormalTok{, }\FloatTok{1.4934}\NormalTok{,}
  \FloatTok{1.4933}\NormalTok{, }\FloatTok{1.4931}\NormalTok{, }\FloatTok{1.4930}\NormalTok{, }\FloatTok{1.4928}\NormalTok{, }\FloatTok{1.4927}\NormalTok{, }\FloatTok{1.4926}\NormalTok{, }\FloatTok{1.4924}\NormalTok{, }\FloatTok{1.4923}\NormalTok{, }\FloatTok{1.4922}\NormalTok{, }\FloatTok{1.4921}\NormalTok{,}
  \FloatTok{1.4920}\NormalTok{, }\FloatTok{1.4918}\NormalTok{, }\FloatTok{1.4917}\NormalTok{, }\FloatTok{1.4916}\NormalTok{, }\FloatTok{1.4915}\NormalTok{, }\FloatTok{1.4914}\NormalTok{, }\FloatTok{1.4913}\NormalTok{, }\FloatTok{1.4912}\NormalTok{, }\FloatTok{1.4911}\NormalTok{, }\FloatTok{1.4910}\NormalTok{)}
\NormalTok{factors.thd.sqrt }\OtherTok{\textless{}{-}} \FunctionTok{c}\NormalTok{(}
      \ConstantTok{NA}\NormalTok{, }\FloatTok{1.7725}\NormalTok{, }\FloatTok{1.6455}\NormalTok{, }\FloatTok{2.0172}\NormalTok{, }\FloatTok{1.6774}\NormalTok{, }\FloatTok{1.6887}\NormalTok{, }\FloatTok{1.6810}\NormalTok{, }\FloatTok{1.6363}\NormalTok{, }\FloatTok{1.6431}\NormalTok{, }\FloatTok{1.6137}\NormalTok{,}
  \FloatTok{1.6036}\NormalTok{, }\FloatTok{1.5938}\NormalTok{, }\FloatTok{1.5826}\NormalTok{, }\FloatTok{1.5771}\NormalTok{, }\FloatTok{1.5683}\NormalTok{, }\FloatTok{1.5639}\NormalTok{, }\FloatTok{1.5574}\NormalTok{, }\FloatTok{1.5530}\NormalTok{, }\FloatTok{1.5488}\NormalTok{, }\FloatTok{1.5449}\NormalTok{,}
  \FloatTok{1.5417}\NormalTok{, }\FloatTok{1.5385}\NormalTok{, }\FloatTok{1.5361}\NormalTok{, }\FloatTok{1.5333}\NormalTok{, }\FloatTok{1.5313}\NormalTok{, }\FloatTok{1.5290}\NormalTok{, }\FloatTok{1.5272}\NormalTok{, }\FloatTok{1.5254}\NormalTok{, }\FloatTok{1.5238}\NormalTok{, }\FloatTok{1.5224}\NormalTok{,}
  \FloatTok{1.5210}\NormalTok{, }\FloatTok{1.5198}\NormalTok{, }\FloatTok{1.5185}\NormalTok{, }\FloatTok{1.5175}\NormalTok{, }\FloatTok{1.5163}\NormalTok{, }\FloatTok{1.5155}\NormalTok{, }\FloatTok{1.5144}\NormalTok{, }\FloatTok{1.5136}\NormalTok{, }\FloatTok{1.5127}\NormalTok{, }\FloatTok{1.5119}\NormalTok{,}
  \FloatTok{1.5111}\NormalTok{, }\FloatTok{1.5104}\NormalTok{, }\FloatTok{1.5097}\NormalTok{, }\FloatTok{1.5091}\NormalTok{, }\FloatTok{1.5085}\NormalTok{, }\FloatTok{1.5078}\NormalTok{, }\FloatTok{1.5073}\NormalTok{, }\FloatTok{1.5067}\NormalTok{, }\FloatTok{1.5063}\NormalTok{, }\FloatTok{1.5057}\NormalTok{,}
  \FloatTok{1.5053}\NormalTok{, }\FloatTok{1.5048}\NormalTok{, }\FloatTok{1.5044}\NormalTok{, }\FloatTok{1.5039}\NormalTok{, }\FloatTok{1.5035}\NormalTok{, }\FloatTok{1.5031}\NormalTok{, }\FloatTok{1.5027}\NormalTok{, }\FloatTok{1.5024}\NormalTok{, }\FloatTok{1.5020}\NormalTok{, }\FloatTok{1.5017}\NormalTok{,}
  \FloatTok{1.5013}\NormalTok{, }\FloatTok{1.5010}\NormalTok{, }\FloatTok{1.5007}\NormalTok{, }\FloatTok{1.5004}\NormalTok{, }\FloatTok{1.5001}\NormalTok{, }\FloatTok{1.4998}\NormalTok{, }\FloatTok{1.4995}\NormalTok{, }\FloatTok{1.4993}\NormalTok{, }\FloatTok{1.4990}\NormalTok{, }\FloatTok{1.4988}\NormalTok{,}
  \FloatTok{1.4986}\NormalTok{, }\FloatTok{1.4983}\NormalTok{, }\FloatTok{1.4981}\NormalTok{, }\FloatTok{1.4979}\NormalTok{, }\FloatTok{1.4977}\NormalTok{, }\FloatTok{1.4974}\NormalTok{, }\FloatTok{1.4972}\NormalTok{, }\FloatTok{1.4970}\NormalTok{, }\FloatTok{1.4969}\NormalTok{, }\FloatTok{1.4966}\NormalTok{,}
  \FloatTok{1.4965}\NormalTok{, }\FloatTok{1.4963}\NormalTok{, }\FloatTok{1.4961}\NormalTok{, }\FloatTok{1.4959}\NormalTok{, }\FloatTok{1.4958}\NormalTok{, }\FloatTok{1.4956}\NormalTok{, }\FloatTok{1.4955}\NormalTok{, }\FloatTok{1.4953}\NormalTok{, }\FloatTok{1.4952}\NormalTok{, }\FloatTok{1.4950}\NormalTok{,}
  \FloatTok{1.4949}\NormalTok{, }\FloatTok{1.4947}\NormalTok{, }\FloatTok{1.4946}\NormalTok{, }\FloatTok{1.4944}\NormalTok{, }\FloatTok{1.4943}\NormalTok{, }\FloatTok{1.4942}\NormalTok{, }\FloatTok{1.4940}\NormalTok{, }\FloatTok{1.4940}\NormalTok{, }\FloatTok{1.4938}\NormalTok{, }\FloatTok{1.4937}\NormalTok{)}

\NormalTok{mad.generic }\OtherTok{\textless{}{-}} \ControlFlowTok{function}\NormalTok{(med, factors, alpha, beta) }\ControlFlowTok{function}\NormalTok{(x) \{}
\NormalTok{  n }\OtherTok{\textless{}{-}} \FunctionTok{length}\NormalTok{(x)}
\NormalTok{  factor }\OtherTok{\textless{}{-}} \FunctionTok{ifelse}\NormalTok{(n }\SpecialCharTok{\textless{}=} \DecValTok{100}\NormalTok{, factors[n], }\DecValTok{1} \SpecialCharTok{/} \FunctionTok{qnorm}\NormalTok{(}\FloatTok{0.75}\NormalTok{) }\SpecialCharTok{/}\NormalTok{ (}\DecValTok{1} \SpecialCharTok{+}\NormalTok{ alpha }\SpecialCharTok{/}\NormalTok{ n }\SpecialCharTok{+}\NormalTok{ beta }\SpecialCharTok{/}\NormalTok{ n}\SpecialCharTok{\^{}}\DecValTok{2}\NormalTok{))}
  \FunctionTok{med}\NormalTok{(}\FunctionTok{abs}\NormalTok{(x }\SpecialCharTok{{-}} \FunctionTok{med}\NormalTok{(x))) }\SpecialCharTok{*}\NormalTok{ factor}
\NormalTok{\}}
\NormalTok{mad.sm }\OtherTok{\textless{}{-}} \FunctionTok{mad.generic}\NormalTok{(med.sm, factors.sm, }\SpecialCharTok{{-}}\FloatTok{0.7668}\NormalTok{, }\SpecialCharTok{{-}}\FloatTok{2.1897}\NormalTok{)}
\NormalTok{mad.hd }\OtherTok{\textless{}{-}} \FunctionTok{mad.generic}\NormalTok{(med.hd, factors.hd, }\SpecialCharTok{{-}}\FloatTok{0.4912}\NormalTok{, }\SpecialCharTok{{-}}\FloatTok{7.6350}\NormalTok{)}
\NormalTok{mad.thd.sqrt }\OtherTok{\textless{}{-}} \FunctionTok{mad.generic}\NormalTok{(med.thd.sqrt, factors.thd.sqrt, }\SpecialCharTok{{-}}\FloatTok{0.6954}\NormalTok{, }\SpecialCharTok{{-}}\FloatTok{4.9261}\NormalTok{)}
\end{Highlighting}
\end{Shaded}

\newpage

\printbibliography

\end{document}